\documentclass[aps,prd,10pt,notitlepage,superscriptaddress,nofootinbib,numbers]{revtex4-2}

\usepackage[utf8]{inputenc}
\usepackage[T1]{fontenc}
\usepackage{lmodern}
\usepackage{microtype}
\usepackage{amsmath,amssymb,amsfonts,mathrsfs}
\usepackage{bm}
\usepackage{graphicx}
\usepackage{float}
\usepackage[dvipsnames]{xcolor}
\usepackage{hyperref}
\hypersetup{colorlinks=true,citecolor=Purple,linkcolor=Purple,urlcolor=Purple,breaklinks=true}
\usepackage[capitalize]{cleveref}
\usepackage{tikz}

\definecolor{lime}{HTML}{A6CE39}
\DeclareRobustCommand{\orcidicon}{%
	\begin{tikzpicture}
		\draw[lime, fill=lime] (0,0) 
		circle [radius=0.16] 
		node[white] {{\fontfamily{qag}\selectfont \tiny ID}};
		\draw[white, fill=white] (-0.0625,0.095) 
		circle [radius=0.007];
	\end{tikzpicture}
	\hspace{-2mm}
}

\foreach \x in {A, ..., Z}{%
	\expandafter\xdef\csname orcid\x\endcsname{\noexpand\href{https://orcid.org/\csname orcidauthor\x\endcsname}{\noexpand\orcidicon}}
}

\newcommand\orcidJonathan{{\href{https://orcid.org/0000-0001-9291-0893}{\orcidicon}}}
\newcommand\orcidEdson{{\href{https://orcid.org/0000-0001-9929-5977}{\orcidicon}}}
\newcommand\orcidAslam{{\href{https://orcid.org/0009-0006-5160-1131}{\orcidicon}}}
\newcommand\orcidSaleem{{\href{https://orcid.org/0000-0002-1124-9624}{\orcidicon}}}

\newcommand{\Del}{\nabla}
\newcommand{\dd}{\mathrm{d}}

\begin{document}

\raggedbottom
\setlength{\textfloatsep}{12pt plus 2pt minus 2pt}
\setlength{\floatsep}{10pt plus 2pt minus 2pt}
\setlength{\intextsep}{10pt plus 2pt minus 2pt}
\setlength{\abovedisplayskip}{7pt plus 2pt minus 2pt}
\setlength{\belowdisplayskip}{7pt plus 2pt minus 2pt}
\setlength{\abovedisplayshortskip}{4pt plus 1pt minus 1pt}
\setlength{\belowdisplayshortskip}{5pt plus 1pt minus 1pt}
\renewcommand{\topfraction}{0.90}
\renewcommand{\bottomfraction}{0.85}
\renewcommand{\textfraction}{0.10}
\renewcommand{\floatpagefraction}{0.70}
\makeatletter
\setlength{\@fptop}{0pt}
\setlength{\@fpsep}{12pt plus 2pt minus 2pt}
\setlength{\@fpbot}{0pt plus 1fil}
\makeatother

\title{Thin-Shell Wormholes in Born--Infeld-Type Magnetic Spacetimes}

\author{Jonathan A. Rebou\c{c}as\orcidJonathan\!\!}
\email{jalvesreboucas@ifce.edu.br}
\affiliation{Instituto Federal de Educa\c{c}\~ao, Ci\^encia e Tecnologia do Cear\'a (IFCE), Iguatu, Brazil}

\author{Edson Otoniel\orcidEdson\!\!}
\email{edson.otoniel@ufca.edu.br}
\affiliation{Universidade Federal do Cariri (UFCA), Instituto de Forma\c{c}\~ao de Educadores - IFE,  R. Oleg\'ario Emidio de Araujo S/N, Brejo Santo - CE, 63.260-000 - Brazil}

\author{Muhammad Israr Aslam\orcidAslam\!\!}
\email{israr.aslam@umt.edu.pk, mrisraraslam@gmail.com (Corresponding author)}
\affiliation{Department of Mathematics, School of Science, University of Management and Technology, Lahore-$54770$, Pakistan}

\author{Rabia Saleem\orcidSaleem\!\!}
\email{rabiasaleem@cuilahore.edu.pk}
\affiliation{Department of Mathematics, COMSATS University Islamabad, Lahore Campus, Lahore-$54000$ Pakistan.}

\date{\today}

\begin{abstract}
We investigate the radial stability, thermodynamic balance, and bilateral disk images of symmetric thin-shell wormholes built from Born--Infeld-type magnetic exteriors. The exact lapse and junction conditions define a fixed-mass family in which magnetic self-energy ties the nonlinear scale to the charge. The shell has negative energy density, although an inner sampled region satisfies the tangential null and intrinsic strong energy conditions. Linear radial perturbations yield unstable calibrated barotropic samples and stable sectors for the generalized and modified cosmic Chaplygin laws on selected real branches. Analytic horizon and large-radius limits explain the stability transitions and show how the modified linear pressure term can replace outer stability with finite radial intervals. Minimally coupled Maxwell rays traverse transparent junctions between exteriors with a common arctangent emission profile, viewed by a static observer. The wormhole-black-hole intensity difference separates transmitted disk light from emission removed by throat excision. Compared with static emitters, circular and plunging motion reduce the face-on intensities in the studied cases; inclined motion generates strong azimuthal contrast. Increasing the subcritical charge contracts the critical circle and permits more disk crossings near it in the optical samples. Under a reversible first law, flux cancellation gives isentropic motion at positive temperature, with entropy dependence left to a constitutive completion. These results distinguish geometric, material, and kinematic effects on shell support and disk illumination.
\end{abstract}

\maketitle
\vspace{-4pt}
\tableofcontents

\section{Introduction}\label{sec:introduction}
The theory of General Relativity (GR) is a highly successful theory for describing gravitational phenomena around astrophysical objects. Among its most remarkable achievements, one can mention the predictions of astrophysical black holes and the detection of gravitational waves. Both phenomena have since been confirmed by astronomical observations, and the measurements coincide with GR predictions within the domains tested \cite{Will2014,thin2,thin3}. However, despite the remarkable success of the theory, GR still presents several theoretical challenges that need to be addressed through effective implementations. Particularly, the Planck scale is expected to play a major role in the development of a quantum theory of gravity that provides a consistent ultraviolet (UV) completion of GR. In the absence of a fundamental quantum theory of gravity, extended theories of gravity have been explored as phenomenological extensions of GR. These mechanisms aim to explain the gravitational phenomena beyond GR, while staying consistent with the recent observational data and constraints imposed by local gravitational tests. In the context of GR coupled to nonlinear electrodynamics, black holes governed by this matter theory have attracted increasing attention, specifically in view of recent advances in astrophysical observations. This interest stems from the fact that astrophysical black holes can be related to intense magnetic fields and are often surrounded by plasma, making nonlinear electrodynamics a useful theoretical platform for discussing their intricate properties and visual signatures. Hence, it is important to investigate the physical properties of black holes sourced by nonlinear electrodynamics \cite{thin4}.

In the context of Maxwell's electrodynamics, a point charge produces a singularity at its location, resulting in infinite electromagnetic self energy. Born and Infeld proposed a nonlinear electromagnetic field with the aim of solving this problem \cite{thin5}. Motivated by this seminal work, various studies have investigated the coupling of GR with the Born-Infeld nonlinear electrodynamics mechanism to investigate the singularity problem in black hole solutions and their relevant dynamics \cite{thin6,thin7,thin8}. Additionally, generalized forms of the Born–Infeld action naturally arise as leading order contributions in effective field theories obtained from superstring theory \cite{thin9,thin10,thin11}. From an astrophysical perspective, nonlinear electrodynamics plays an important role in describing the physical properties of highly magnetized compact objects, i.e., pulsars and magnetars, where electromagnetic fields can be extremely strong \cite{thin12}. The regular black hole model introduced by Bardeen was subsequently interpreted as a nonlinear magnetic monopole by Ay\'on-Beato and Garc\'ia \cite{AyonBeatoGarcia2000}. Consequently, various publications have been devoted to exploring the physical properties of black holes sourced by nonlinear electrodynamics; for instance, see Refs. \cite{thin14,thin15,thin16}. Moreover, with the development of theoretical modeling, the study of black hole shadows, light deflection angles and dynamics of photon rings has become a powerful diagnostic tool for testing nonlinear electrodynamics and related compact object models in strong gravitational fields \cite{thin17,thin18,thin19,thin20}. Related calculations have examined magnetically charged Born--Infeld black hole images \cite{WenHongTao2023} and static disks with an arctangent emissivity \cite{ChenYang2025}.

Traversable wormholes are hypothetical structures in spacetime that represent tunnel connections between two distinct parts of spacetime or may provide a path between two different universes. These intriguing configurations can emerge as solutions to Einstein's GR and are characterized by a nontrivial spacetime topology that distinguishes them from ordinary flat spacetime. They describe horizonless spacetime geometries, featuring a minimal surface area that links two distinct spacetime regions, which is well known as the wormhole throat. The idea of wormholes was developed by Wheeler within the fabric of spacetime quantum foam \cite{thin21}, while the first mathematical description of such structures can be traced back to the Einstein-Rosen bridge, which was introduced in 1935 \cite{thin22}. The latter is not traversable. However, the idea of modern traversable wormholes came from the pioneering work of Morris and Thorne \cite{thin23}, who showed that a traversable wormhole geometry may exist under certain physical constraints. Further, the scalar field configurations of Ellis and Bronnikov provided early explicit realizations of regular structures with nontrivial topology \cite{thin24,thin25}. These theoretical constructions have established wormholes as valuable frameworks for exploring the limitations of GR and examining whether effective models or quantum inspired corrections can support spacetime geometries that are difficult to realize using ordinary matter \cite{thin26,thin27}. Moreover, considerable attention has been devoted to identifying their potential astrophysical signatures and exploring observational methods for discovering such structures \cite{thin28,thin29}.

In the context of traversable wormhole configurations, the most important attribute is the flare out condition, which is determined by the geometry of the wormhole throat. To support the throat within GR, violations of the usual null energy condition are required \cite{thin27}. This requirement does not by itself ensure stability against radial perturbations. This feature has motivated inquiries into mechanisms that localize the exotic matter to a narrow region, minimize its overall contribution, or reinterpret its effects as effective sources arising from additional gravitational degrees of freedom. Since exotic matter is hypothetical and possesses unusual physical properties, minimizing its required amount remains a significant challenge in the study of wormhole dynamics. To overcome this problem, various wormhole solutions have been explored through effective implementations. In the context of modified theories of gravity, the study of wormhole solutions becomes increasingly intriguing due to the possibility of obtaining feasible solutions beyond the constraints of GR \cite{thin30,thin31,thin32,thin33,thin34,thin35,thin36,thin37,thin38}. Consequently, examples include wormholes supported by phantom sectors \cite{thin39,thin40} and constructions based on quantum vacuum effects \cite{thin33}; wormhole configurations in de Sitter space are also found in the literature \cite{thin41}. In this perspective, wormhole research extends beyond the exploration of exotic solutions, offering a significant framework for investigating how microscopic, effective, or emergent gravitational phenomena can shape the permissible macroscopic topology of spacetime.

Timelike thin shell geometries are an important class of gravitational configurations in which two distinct spacetime regions are connected across a hypersurface carrying a surface stress energy tensor. The geometry of these configurations is commonly developed using the Darmois--Israel junction conditions, which establish the framework for joining two spacetime manifolds across a hypersurface. This formalism enables the determination of the extrinsic curvature on either side of the shell and the corresponding stress energy tensor \cite{thin42,thin43,Israel1966}. These shells are commonly obtained using the cut and paste construction, so that the matter required to support the throat is confined to an infinitesimally thin layer rather than spread throughout the bulk \cite{thin45,PoissonVisser1995}. This approach separates the bulk spacetime geometry on either side of the shell from the matter confined to its hypersurface, thereby providing a systematic way to examine directly how a given seed spacetime influences the surface energy conditions and the throat dynamics.

Although a physically admissible timelike junction can be developed, its presence alone does not ensure dynamical stability. The linear stability technique developed in Ref.~\cite{PoissonVisser1995} shows that the shell's response to radial perturbations depends on both the underlying spacetime geometry and the equation of state governing the matter supporting the shell. Later investigations extended this stability treatment to encompass general spherically symmetric thin shell configurations \cite{thin46,thin47}, including cosmological, charged, dilatonic, and generalized Chaplygin gas frameworks \cite{thin48,thin49,thin50,thin51,thin52,thin53}. Recently, alternative thermodynamic approaches have also been considered for assessing the stability of Schwarzschild thin shell wormholes \cite{thin54}, and the combined effects of dynamical and thermodynamic stability have been studied in charged thin shell configurations \cite{thin56}. Radial stability has also been investigated for wormhole configurations constructed from quantum corrected polymer black holes \cite{thin55}. The shell first law relates reversible entropy evolution to the bulk flux \cite{LoboRodriguesThermodynamics2026}, whereas acceleration and particle creation temperatures require separate definitions \cite{LoboRodriguesUnruh2026}. These developments indicate that the viability of thin shell wormholes is governed by several factors, such as the matter properties encoded in the shell equation of state, the horizon structure, and the local behavior of the bulk metric function and its associated consequences. More recently, Rebou\c{c}as and collaborators have explored the intricate properties of thin shell wormholes through effective implementations \cite{jn1,jn2,jn3}. Specifically, they examined entropy induced black hole geometries, junction conditions, stability, thermodynamic properties, and the visual characteristics of reflection asymmetric polymer thin shell wormholes surrounded by Kiselev type quintessence on one side. Related studies have also considered symmetric shells constructed from a regularized string cloud geometry \cite{ZhongEtAl2026RegularizedShell}.

Born--Infeld thin shell wormholes have previously been studied with a generalized Chaplygin gas \cite{EiroaFigueroaAguirre2012BI} and with the generalized cosmic Chaplygin gas \cite{Azam2016BIGCCG}. The cosmic law introduced by Gonz\'alez-D\'iaz \cite{GonzalezDiaz2003GCCG}, its applications to shell stability \cite{SharifMumtaz2014CosmicChaplygin}, and the modified cosmic extension used in cosmological fluid models \cite{SadeghiFarahani2013MCCG} provide the constitutive antecedents relevant here. These developments motivate examining the magnetic family in which self energy supplies the mass and links the charge to the nonlinear scale. At fixed mass, this constraint defines a specific comparison between exterior geometries and their shell responses, while the nonlinear exponent includes standard Born--Infeld electrodynamics as a particular case. Our contribution combines that constrained magnetic family with calibrated barotropic and cosmic Chaplygin responses and analytic horizon and large radius limits. The accompanying bilateral disk comparison distinguishes light transmitted through the throat from emission removed by excision, and compares static and moving emitters to identify the influence of source kinematics.

Guided by insights from prior research, this work aims to explore the properties of thin shell wormhole solutions based on Born--Infeld type magnetic exteriors. Starting from the static, spherically symmetric magnetic solution of the Born--Infeld type nonlinear electrodynamics model as studied in \cite{Kruglov2017,KruglovCorrigendum}, we use the corresponding metric function and develop the wormhole by gluing two exterior copies of the corresponding spacetime. We then derive the expressions for the shell stresses and energy conditions and discuss the radial stability using the effective potential. We evaluate three constitutive laws for the surface matter, namely the linear barotropic, generalized cosmic Chaplygin and modified cosmic Chaplygin laws, and use the equilibrium criterion to fix one constitutive parameter for each static shell. We also discuss the surface stresses, which determine the mechanical energy of the layer, and a thermodynamic interpretation through the shell first law in the magnetic geometry. Finally, we discuss the visual characteristics of the thin shell wormholes through intensity maps of bilateral disks, comparing static and moving emitters viewed by a static observer. We use minimally coupled Maxwell test rays and geometrically selected throats for this optical comparison.

This paper is organized as follows. In Sect. \ref{sec:seed}, we will define the background of the static, spherically symmetric magnetic solution of the Born--Infeld type nonlinear electrodynamics model and its causal domains. In Sect. \ref{sec:construction}, we develop the symmetric thin shell wormhole and derive the induced metric, proper time normalization, unit normal, and extrinsic curvature components along with energy conditions. In Sect. \ref{sec:general_stability}, we illustrate the conservation equation and the radial effective potential. In Sect. \ref{sec:eos_stability}, we introduce the linear barotropic, generalized cosmic Chaplygin and modified cosmic Chaplygin shell models and discuss the stability analysis. In Sect. \ref{sec:thermodynamics}, we will discuss the thermodynamics of the magnetic throat in depth. In Sect. \ref{sec:optical_appearance}, we will discuss the optical signatures of the bilateral disk sources through the intensity models. Finally, the last section will conclude the whole study by revisiting the main outcomes.

\section{Magnetic seed geometry and admissible exterior}\label{sec:seed}

We take as the bulk geometry the static, spherically symmetric magnetic solution of the Born-Infeld-type nonlinear electrodynamics model studied by Kruglov \cite{Kruglov2017,KruglovCorrigendum}. In Schwarzschild gauge, the line element used below is
\begin{equation}
\dd s^2=-\mathcal{F}(r)\dd t^2+\frac{\dd r^2}{\mathcal{F}(r)}+r^2\left(\dd\theta^2+\sin^2\!\theta\,\dd\varphi^2\right),
\label{eq:seed_metric}
\end{equation}
where $r$ is the areal radius and the lapse is
\begin{equation}
\mathcal{F}(r)=1-\frac{2Gm_{M}}{r}-\frac{2Gr^2}{3\beta}
\left[{}_2F_1\!\left(-\frac34,-\nu;\frac14;-\frac{\beta q^2}{2\nu r^4}\right)-1\right].
\label{eq:seed_lapse}
\end{equation}
Here $G$ is Newton's constant, $m_M$ is the magnetic mass parameter, $q$ is the magnetic charge, $\beta$ has dimensions of length to the fourth power in Kruglov's electromagnetic units, and $\nu$ is dimensionless. We write the seed exponent as $\nu$ (denoted by $\sigma$ in Kruglov's notation) to distinguish it from the surface density $\sigma$ of the shell. The symbol ${}_2F_1$ denotes the Gauss hypergeometric function. A prime on $\mathcal{F}$ denotes differentiation with respect to $r$.

The lapse in \cref{eq:seed_lapse} follows from the nonlinear electromagnetic Lagrangian
\begin{equation}
\mathscr{L}=\frac{1}{\beta}\left[1-\left(1+\frac{\beta\mathscr{I}}{\nu}-\frac{\beta\gamma\mathscr{J}^{2}}{2\nu}\right)^{\!\nu}\right],
\label{eq:nled_lagrangian}
\end{equation}
where $\mathscr{I}=F_{\mu\nu}F^{\mu\nu}/4$, $\mathscr{J}=F_{\mu\nu}\widetilde F^{\mu\nu}/4$, and $\gamma$ is another length-four parameter of the full theory. The purely magnetic monopole has $\mathscr{J}=0$ and $\mathscr{I}=q^2/(2r^4)$, so the metric used in this paper is independent of $\gamma$. Consequently, a change in $\gamma$ cannot generate a new purely magnetic thin-shell geometry within the branch considered here.

The choice $\nu=1/2$ with $\beta=\gamma$ recovers standard Born--Infeld electrodynamics, while the weak-field expansion approaches Maxwell theory. At large $r$, the lapse has the Reissner--Nordstr\"om form at leading orders,
\begin{equation}
\mathcal{F}(r)=1-\frac{2Gm_M}{r}+\frac{Gq^2}{r^2}+\mathcal{O}(r^{-6}),
\label{eq:rn_asymptotic}
\end{equation}
with the first nonlinear correction controlled by $\beta(1-\nu)q^4/(\nu r^6)$. Equation \eqref{eq:rn_asymptotic} is used only as an asymptotic check; all shell expressions below retain the exact lapse.

The nonlinear electromagnetic model admits $0\leq\nu<1$ with nonnegative nonlinear parameters. We take $\beta>0$ and $q\neq0$ and restrict the finite magnetic self-energy branch to
\begin{equation}
0<\nu<\frac34,
\label{eq:seed_domain}
\end{equation}
Within this interval, the magnetic contribution accounts for the entire asymptotic mass, with no independent bare-mass term. The mass constraint is
\begin{equation}
m_M=\frac{\mathcal C(\nu)|q|^{3/2}}{(2\nu)^{3/4}\beta^{1/4}},\qquad
\mathcal C(\nu)=\frac{\nu}{3}\operatorname{B}\!\left(\frac14,\frac34-\nu\right).
\label{eq:magnetic_self_mass}
\end{equation}
Here $\mathcal C(\nu)$ is a dimensionless normalization function and $\operatorname B$ is the Euler beta function. In the electromagnetic units of the lapse, $m_M$ has inverse-length dimensions. The geometric mass is $M=Gm_M>0$, with dimensions of length. We introduce the corresponding geometric magnetic parameters through
\begin{equation}
Q=\sqrt{G}\,q,\qquad \beta_{\rm g}=\frac{\beta}{G}.
\label{eq:geometric_magnetic_parameters}
\end{equation}
The geometric charge $Q$ has dimensions of length and $\beta_{\rm g}$ has dimensions of length squared. The mass constraint then becomes
\begin{equation}
\frac{\beta_{\rm g}}{M^2}=\frac{\mathcal C(\nu)^4}{(2\nu)^3}\left(\frac{|Q|}{M}\right)^6.
\label{eq:mass_normalized_magnetic_constraint}
\end{equation}
At fixed $M$ and $\nu$, the nonlinear scale varies as the sixth power of the charge. Thus a charge variation in this family changes $\beta_{\rm g}$ consistently with the electromagnetic origin of the mass.

Throughout the geometric description we use $G=c=k_B=1$ and retain $\hbar$, whose dimensions are length squared. The explicit factors of $G$ in the seed relations keep track of the conversion from electromagnetic to geometric quantities. Surface density and pressure have inverse-length dimensions, geometric energies and temperatures have length dimensions, and entropy is dimensionless. The geometric mass supplies the common reference length: numerical evaluations set $M=1$, while analytic expressions retain it explicitly. Each dimensionful geometric quantity is normalized by the corresponding power of $M$; lengths and areas are divided by $M$ and $M^2$, surface stresses are multiplied by $M$, and radial-potential curvatures by $M^2$. In this convention, the numerical lapse inputs $q$ and $\beta$ coincide with $Q/M$ and $\beta_{\rm g}/M^2$, respectively.

Horizons are positive roots of $\mathcal F(r)=0$; we denote the largest positive root by $r_h$. Finite self-energy does not ensure that such a root exists. For $0<\nu<1/2$, the horizon-disappearance threshold in the constrained fixed-mass family is determined by
\begin{equation}
\mathcal F(r_e;Q_c)=0,\qquad \mathcal F'(r_e;Q_c)=0.
\label{eq:seed_critical_charge}
\end{equation}
Here $Q_c=Q_c(\nu;M)>0$ is the critical geometric magnetic charge and $r_e>0$ is the degenerate-horizon radius; both have dimensions of length. The lapse and its radial derivative are evaluated at fixed $M$ and $\nu$, with the nonlinear scale constrained by \cref{eq:mass_normalized_magnetic_constraint}. For the positive-charge representative, $Q/Q_c$ is dimensionless and identifies the horizon sector: below unity the seed has horizons, at unity it has a positive-radius degenerate horizon, and above unity it is horizonless. This normalization measures proximity to horizon disappearance for each electromagnetic exponent, making the relevant charge threshold explicit when comparing different members of the magnetic family. It describes a fixed-mass variation, not a variation at fixed nonlinear scale.

At $\nu=1/2$, the threshold is $Q_c/M=\sqrt{2}/\mathcal C(1/2)$, but the limiting radius is zero, so a positive-radius horizon requires $0<Q/Q_c<1$. For $1/2<\nu<3/4$, the central lapse tends to negative infinity and the asymptotic lapse tends to unity; continuity therefore guarantees at least one positive horizon for every nonzero charge on this mass branch. No finite horizon-disappearance charge exists in that sector, where $Q/M$ supplies the dimensionless charge. We retain horizon-bearing seeds: a strictly subcritical charge suffices throughout $0<\nu\leq1/2$, while the degenerate boundary below $\nu=1/2$ can be treated separately. The positive roots are generally determined numerically.

The retained exterior begins outside the outer seed horizon, and the lapse remains positive throughout the shell's allowed motion. For $0<\nu<1/2$, causality and unitarity of the nonlinear electromagnetic theory additionally require
\begin{equation}
 r^4>\frac{\beta q^2(1-2\nu)}{2\nu}.
\label{eq:field_bound}
\end{equation}
This restriction determines the admissible bulk region independently of the surface equation of state. The central curvature invariant diverges even when the lapse has a finite limiting value. Excision removes that singularity and the seed horizon from the joined manifold without making the original magnetic spacetime globally regular.

\section{Symmetric cut-and-paste construction}\label{sec:construction}

Within the seed sector defined in \cref{sec:seed}, we choose a radius $a>r_h$ in the outer static component, with $\mathcal F(a)>0$ and, where applicable, \cref{eq:field_bound}. We remove the region with $r<a$ from two identical copies of the seed exterior and retain the two manifolds with boundary $r=a$. The boundaries are identified to form a timelike throat $\Sigma$; the areal radius increases away from $\Sigma$ on either side, so the identified surface is a minimal-area sphere in the glued geometry.

The throat is allowed to move radially and is described intrinsically by coordinates $\xi^i=(\tau,\theta,\varphi)$, where $\tau$ is proper time along the shell. A convenient embedding into either copy is written componentwise as
\begin{equation}
X^\mu_{\pm}(\xi^i)=\bigl(t_{\pm}(\tau),a(\tau),\theta,\varphi\bigr).
\label{eq:embedding}
\end{equation}
The normalization of the shell trajectory with respect to \cref{eq:seed_metric} determines $\dot t_{\pm}$; a dot denotes differentiation with respect to $\tau$. Pulling back the bulk metric gives the first fundamental form,
\begin{equation}
h_{ij}=g_{\mu\nu}\frac{\partial X^\mu}{\partial\xi^i}\frac{\partial X^\nu}{\partial\xi^j},
\label{eq:pullback}
\end{equation}
where $h_{ij}$ is the induced metric, $g_{\mu\nu}$ is the bulk metric, $X^\mu$ is the embedding, and Greek and Latin indices refer to bulk and shell coordinates, respectively. Substitution of \cref{eq:embedding} and the proper-time normalization yields
\begin{equation}
\dd s^2_{\Sigma}=-\dd\tau^2+a(\tau)^2\left(\dd\theta^2+\sin^2\!\theta\,\dd\varphi^2\right).
\label{eq:induced_metric}
\end{equation}

Let $\Phi(r,\tau)=r-a(\tau)$ define the shell as a level surface. The two normals are oriented toward increasing $r$ in their respective retained regions and can be written as the normalized gradient together with their explicit form,
\begin{equation}
n^{\pm}_{\mu}=\pm\left|g^{\alpha\beta}\partial_{\alpha}\Phi\,\partial_{\beta}\Phi\right|^{-1/2}\partial_{\mu}\Phi
=\pm\left(-\dot a,\frac{\sqrt{\mathcal{F}(a)+\dot a^2}}{\mathcal{F}(a)},0,0\right).
\label{eq:unit_normal}
\end{equation}
The normal is orthogonal to the shell trajectory and has unit norm. The $+$ and $-$ orientations are needed because the radial coordinate points away from the throat in opposite directions after gluing.

The extrinsic curvature is defined using the bulk covariant derivative,
\begin{equation}
K^{\pm}_{ij}=-n^{\pm}_{\mu}\left(\frac{\partial^2X^\mu_{\pm}}{\partial\xi^i\partial\xi^j}+\Gamma^{\mu}_{\alpha\beta}\frac{\partial X^\alpha_{\pm}}{\partial\xi^i}\frac{\partial X^\beta_{\pm}}{\partial\xi^j}\right).
\label{eq:extrinsic_definition}
\end{equation}
In the shell's orthonormal basis, the independent components are
\begin{equation}
K^{\hat\theta\,\pm}_{\ \hat\theta}=K^{\hat\varphi\,\pm}_{\ \hat\varphi}=\pm\frac{\sqrt{\mathcal{F}(a)+\dot a^2}}{a},
\label{eq:Kangular}
\end{equation}
\begin{equation}
K^{\hat\tau\,\pm}_{\ \hat\tau}=\pm\frac{\mathcal{F}'(a)+2\ddot a}{2\sqrt{\mathcal{F}(a)+\dot a^2}}.
\label{eq:Ktime}
\end{equation}

For any shell quantity $N$, the jump is $[N]=N^+-N^-$. With the orthonormal surface stress tensor defined by
\begin{equation}
S^{\hat i}{}_{\hat j}=\operatorname{diag}(-\sigma,p,p),
\label{eq:surface_tensor}
\end{equation}
the Israel--Lanczos condition \cite{Israel1966} is
\begin{equation}
S^{\hat i}{}_{\hat j}=-\frac{1}{8\pi}\left([K^{\hat i}{}_{\hat j}]-\delta^{\hat i}{}_{\hat j}[K]\right),
\label{eq:lanczos}
\end{equation}
where $[K]$ is the trace of the jump in the same orthonormal basis. The resulting dynamical surface density and tangential pressure are
\begin{equation}
\sigma(a,\dot a)=-\frac{1}{2\pi a}\sqrt{\mathcal{F}(a)+\dot a^2},
\label{eq:sigma_dynamic}
\end{equation}
\begin{equation}
p(a,\dot a,\ddot a)=\frac{\mathcal{F}'(a)+2\ddot a}{8\pi\sqrt{\mathcal{F}(a)+\dot a^2}}+\frac{\sqrt{\mathcal{F}(a)+\dot a^2}}{4\pi a}.
\label{eq:p_dynamic}
\end{equation}
The negative sign in \cref{eq:sigma_dynamic} is the direct consequence of the opposite outward normals and the angular curvature jump. It represents matter localized on the throat and does not imply that the nonlinear electromagnetic bulk source itself has a negative energy density.

For a static configuration at $a=a_0$ with $\mathcal{F}_0\equiv\mathcal{F}(a_0)>0$, equations \eqref{eq:sigma_dynamic} and \eqref{eq:p_dynamic} reduce to
\begin{equation}
\sigma_0=-\frac{\sqrt{\mathcal{F}_0}}{2\pi a_0},
\label{eq:sigma_static}
\end{equation}
\begin{equation}
p_0=\frac{2\mathcal{F}_0+a_0\mathcal{F}'_0}{8\pi a_0\sqrt{\mathcal{F}_0}},
\label{eq:p_static}
\end{equation}
where $\mathcal{F}'_0\equiv\mathcal{F}'(a_0)$. The static energy-condition combinations used below are
\begin{equation}
\sigma_0+p_0=\frac{a_0\mathcal{F}'_0-2\mathcal{F}_0}{8\pi a_0\sqrt{\mathcal{F}_0}},
\label{eq:nec_combination}
\end{equation}
\begin{equation}
\sigma_0+2p_0=\frac{\mathcal{F}'_0}{4\pi\sqrt{\mathcal{F}_0}}.
\label{eq:sec_combination}
\end{equation}
For the intrinsic $(2+1)$-dimensional surface tensor, the tangential null energy condition (NEC) requires $\sigma+p\geq0$, the weak energy condition (WEC) also requires $\sigma\geq0$, and the dominant energy condition (DEC) requires $\sigma\geq|p|$. Its strong energy condition (SEC) tests $(S_{ij}-S h_{ij})u^iu^j\geq0$ for every timelike vector $u^i$ tangent to $\Sigma$, where $S=-\sigma+2p$; equivalently, it requires $p\geq0$ and $\sigma+p\geq0$. Since $\sigma_0<0$ throughout the retained static region, the WEC and DEC fail. In this sector the tangential NEC already implies $p_0>0$ and $\sigma_0+2p_0>0$, so \cref{eq:sec_combination} is a useful algebraic diagnostic but not an independent intrinsic SEC criterion. Whether the tangential NEC, and hence the intrinsic SEC, holds at a given admissible throat is determined by \cref{eq:nec_combination}; asymptotic flatness alone does not fix its sign.

These intrinsic conditions must be distinguished from the energy conditions on the distributional stress tensor in the four-dimensional spacetime. A null vector crossing the shell normally has a surface NEC contraction proportional to $\sigma_0<0$. Likewise, for a unit timelike observer boosted normally with rapidity $\eta$, the singular coefficient of the four-dimensional SEC contraction is $(\sigma_0+2p_0)/2+\sigma_0\sinh^2\eta$, which becomes negative at sufficiently large $\eta$. Thus neither the full spacetime NEC nor the full spacetime SEC is satisfied by the shell, even where its tangential or intrinsic tests hold.

\begin{figure*}[!htbp]
\centering
\includegraphics[width=\textwidth]{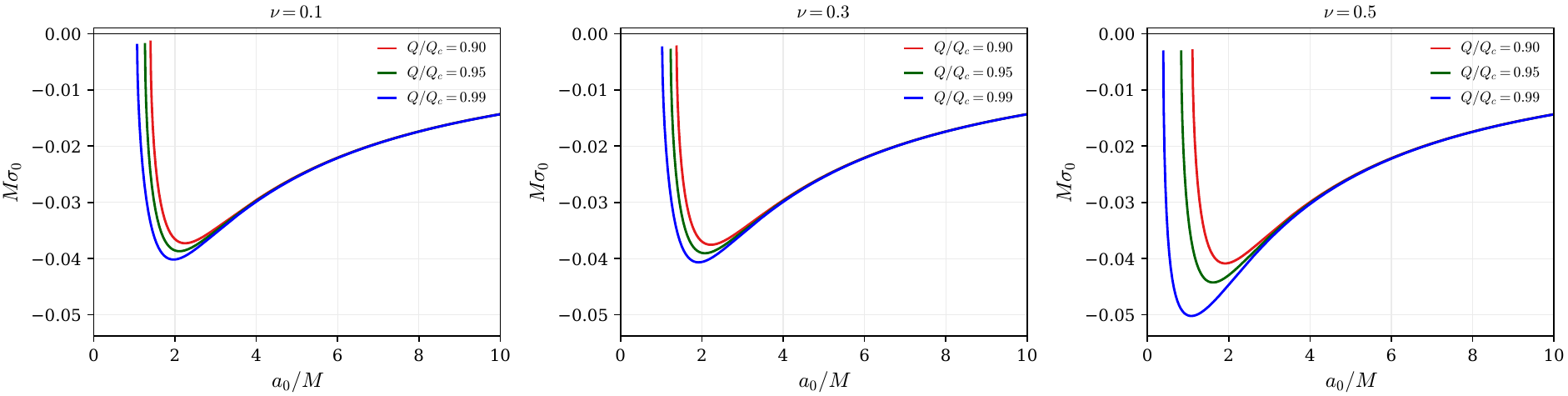}
\caption{Static surface density $M\sigma_0$ as a function of the dimensionless throat radius $a_0/M$. From left to right, the panels correspond to $\nu=0.1$, $0.3$, and $0.5$. The mass is held fixed, and the red, green, and blue curves represent $Q/Q_c=0.90$, $0.95$, and $0.99$, respectively. The nonlinear scale follows the electromagnetic mass constraint in \cref{eq:mass_normalized_magnetic_constraint}. Each curve is restricted to the exterior of the outer seed horizon and, where applicable, to the field domain in \cref{eq:field_bound}. The horizontal black line marks zero density.}
\label{fig:density_selected_nearcritical}
\end{figure*}

The negative-density profiles in Fig.~\ref{fig:density_selected_nearcritical} are nonmonotonic: each sequence reaches a single minimum and then rises toward zero as the static throat radius increases. In Eq.~\eqref{eq:sigma_static}, the increasing exterior lapse competes with the inverse-radius factor, producing the turnover. Differentiating this expression with respect to the static radius shows that a density extremum coincides with saturation of the tangential NEC in Eq.~\eqref{eq:nec_combination}. The minimum therefore identifies an energy-condition boundary along a sequence of distinct static junctions.

For the configurations shown, increasing $Q/Q_c$ shifts the minimum to smaller radii and makes its value more negative. This response is strongest in the Born--Infeld panel, $\nu=0.5$: between $Q/Q_c=0.90$ and $0.99$, the minimum moves from $a_0/M\simeq1.915$ to $1.082$, while $M\sigma_0$ decreases from approximately $-0.0409$ to $-0.0502$. The inward displacement accompanies the shrinking outer seed horizon, which changes the lower boundary of the static throat domain.

The nearly overlapping profiles for $\nu=0.1$ and $0.3$ indicate weak sensitivity within this subcritical family. Comparing corresponding curves at equal $Q/Q_c$, the relative difference in density, measured against the magnitude of the $\nu=0.1$ curve, remains below approximately $1.15\%$ for $2\leq a_0/M\leq10$ and $0.30\%$ for $3\leq a_0/M\leq10$. These values quantify the visual similarity without implying exact independence from $\nu$. At the same normalized charge, the geometric charges in the two panels differ by only about $1.18\%$. Their common mass and closely spaced charges give similar leading exterior terms in Eq.~\eqref{eq:rn_asymptotic}, while the nonlinear corrections decrease rapidly with radius. Differences remain near the inner boundary: for $Q/Q_c=0.99$, the outer horizons occur at $r_h/M\simeq1.0691$ and $1.0237$, respectively.

At large radii, Eqs.~\eqref{eq:sigma_static} and \eqref{eq:rn_asymptotic} imply an inverse-radius decay of the density magnitude, consistent with the merging tails. At the opposite end, the density approaches zero from below as the throat approaches any of these simple outer horizons. This limit does not describe a stress-free shell, since the pressure in Eq.~\eqref{eq:p_static} diverges there; the horizon itself is excluded from the timelike construction. Mechanical admissibility under a specified EOS and radial stability require their separate constitutive and perturbative analyses.

\begin{figure*}[!htbp]
\centering
\includegraphics[width=\textwidth]{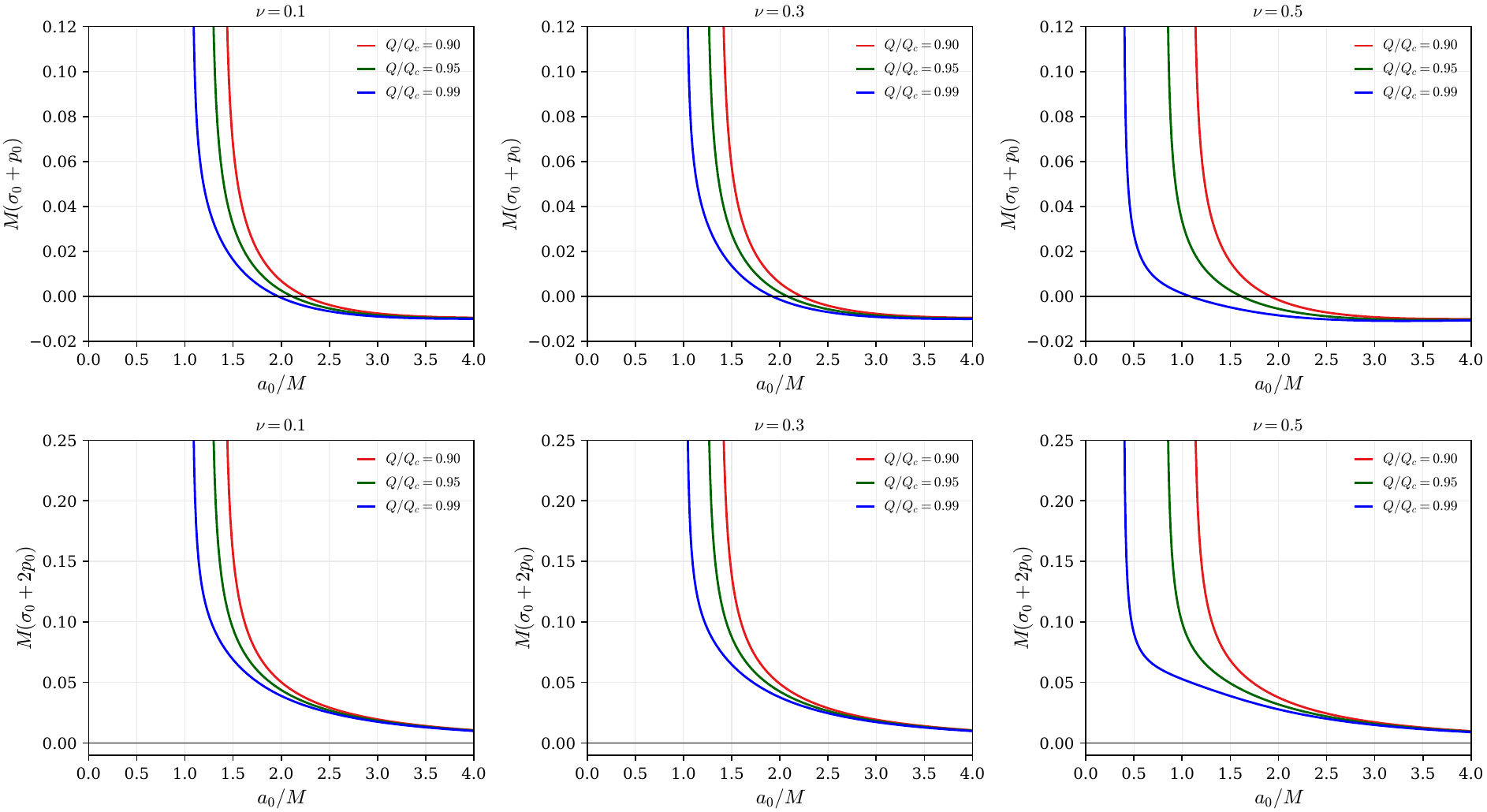}
\caption{Static shell combinations as functions of $a_0/M$ for the same fixed-mass family as Fig.~\ref{fig:density_selected_nearcritical}. The upper row shows $M(\sigma_0+p_0)$ and the lower row shows $M(\sigma_0+2p_0)$; columns correspond to $\nu=0.1$, $0.3$, and $0.5$. Red, green, and blue curves denote $Q/Q_c=0.90$, $0.95$, and $0.99$, respectively. Only radii outside the outer seed horizon and within the magnetic-field domain are retained. The black line marks zero; the positive near-horizon divergence extends above the displayed ordinate range. The lower-row quantity is an algebraic diagnostic, not an independent intrinsic SEC criterion.}
\label{fig:energy_subcritical_2x3}
\end{figure*}

Figure~\ref{fig:energy_subcritical_2x3} resolves the sign change associated with the density minima in Fig.~\ref{fig:density_selected_nearcritical}. Each upper-row curve is positive immediately outside its simple outer horizon, crosses zero once within the plotted interval, and is negative at larger displayed radii. The zeros satisfy Eq.~\eqref{eq:nec_combination} and coincide with the density minima. Increasing $Q/Q_c$ moves this boundary inward in every column; for $\nu=0.5$, it moves from $a_0/M\simeq1.915$ at $Q/Q_c=0.90$ to $1.082$ at $0.99$. Since $\sigma_0<0$, the region between the horizon and each zero satisfies the tangential NEC and the intrinsic surface SEC, whereas both fail beyond the zero. The WEC and DEC, as well as the four-dimensional distributional NEC and SEC tested in normal directions, fail throughout the retained static domain.

The lower-row curves remain positive in the displayed domain. By Eq.~\eqref{eq:sec_combination}, this sign reflects a positive derivative of the lapse and a pressure contribution large enough to outweigh the negative density in $\sigma_0+2p_0$. Positivity of this combination persists beyond the upper-row zero, so it cannot be used by itself to identify the intrinsic SEC region. In particular, the ordinate in the lower row is neither the trace of the surface stress tensor nor an independent energy-condition test.

The first two columns nearly overlap at equal $Q/Q_c$, especially away from their inner boundaries: for $2.5\leq a_0/M\leq4$, corresponding curves differ by less than $3.6\times10^{-4}$ in $M(\sigma_0+p_0)$ and $5.1\times10^{-4}$ in $M(\sigma_0+2p_0)$. Their common mass and closely spaced geometric charges make the leading Reissner--Nordstr\"om terms in Eq.~\eqref{eq:rn_asymptotic} similar, while the nonlinear correction decreases rapidly with radius. The curves and their horizon positions remain distinct near the inner boundary. The $\nu=0.5$ column has a more pronounced inward displacement of the zeros and a different near-horizon profile, particularly at $Q/Q_c=0.99$.

The limiting behavior follows directly from Eqs.~\eqref{eq:sigma_static}--\eqref{eq:sec_combination}. As an admissible throat approaches a simple outer seed horizon from above, $\sigma_0$ tends to zero from below, but $p_0$ and both plotted combinations diverge positively as $(a_0-r_h)^{-1/2}$; the lower combination is twice the upper one at leading order. Thus the clipped upper ends of the curves do not indicate a finite stress at the excluded horizon. At large radius, Eq.~\eqref{eq:rn_asymptotic} gives $\sigma_0+p_0\sim-1/(4\pi a_0)$ and $\sigma_0+2p_0\sim M/(2\pi a_0^2)$. The leading inverse-radius terms of density and pressure cancel only in the latter combination, so the two rows approach zero from opposite sides and at different rates. These radius scans compare distinct static junctions; they do not describe radial evolution or establish mechanical stability.

\section{Conservation law and general radial stability}\label{sec:general_stability}

The intrinsic conservation equation is obtained from the contracted junction identity. For the magnetic seed, the static spherical bulk stress tensor satisfies $T^{t}{}_{t}=T^{r}{}_{r}$, so the normal--tangential flux term vanishes when the two identical sides are matched with the same magnetic branch. The shell therefore obeys
\begin{equation}
\Del_i S^{i}{}_{j}=0.
\label{eq:surface_conservation_covariant}
\end{equation}
For the induced metric \eqref{eq:induced_metric}, the nontrivial radial component becomes
\begin{equation}
\frac{\dd\sigma}{\dd a}=-\frac{2}{a}(\sigma+p).
\label{eq:conservation_radial}
\end{equation}
This equation expresses the balance between the variation of the shell's internal energy and the work done by its tangential pressure as the area changes.

Eliminating $\dot a$ from \cref{eq:sigma_dynamic} gives the one-dimensional equation of motion
\begin{equation}
\dot a^2+V(a)=0,
\label{eq:motion}
\end{equation}
with effective potential
\begin{equation}
V(a)=\mathcal{F}(a)-4\pi^2a^2\sigma(a)^2.
\label{eq:potential}
\end{equation}
The static conditions are $V(a_0)=0$ and $V'(a_0)=0$. The first follows from \cref{eq:sigma_static}; the second follows from \cref{eq:p_static} and the conservation law, so both conditions are imposed before any equation of state is calibrated.

For a general constitutive relation $p=p(\sigma,a)$, define
\begin{equation}
\zeta(a)=\left(\frac{\partial p}{\partial\sigma}\right)_{a},
\label{eq:zeta_definition}
\end{equation}
\begin{equation}
p_a(a)=\left(\frac{\partial p}{\partial a}\right)_{\sigma}.
\label{eq:pa_definition}
\end{equation}
Using \cref{eq:conservation_radial} in two independent differentiations of \cref{eq:potential} gives
\begin{equation}
V'(a)=\mathcal{F}'(a)+8\pi^2a\sigma(a)\bigl[\sigma(a)+2p(a)\bigr],
\label{eq:potential_first_derivative}
\end{equation}
\begin{equation}
V''(a)=\mathcal{F}''(a)-8\pi^2\left\{[\sigma+2p]^2+2\sigma(\sigma+p)(1+2\zeta)-2a\sigma p_a\right\}.
\label{eq:potential_second_derivative}
\end{equation}
At equilibrium, the same result can be written directly in terms of the lapse and the constitutive derivatives evaluated at $(\sigma_0,a_0)$:
\begin{equation}
V''_0=\mathcal{F}''_0-\frac{(\mathcal{F}'_0)^2}{2\mathcal{F}_0}+\frac{\mathcal{F}'_0}{a_0}-\frac{2\mathcal{F}_0}{a_0^2}
 +\frac{2\zeta_0}{a_0}\left(\mathcal{F}'_0-\frac{2\mathcal{F}_0}{a_0}\right)-8\pi\sqrt{\mathcal{F}_0}\,p_{a0}.
\label{eq:potential_second_static}
\end{equation}
The compact dependence on $\mathcal{F}'$ and $\mathcal{F}''$ allows the stability criterion to be evaluated from the exact hypergeometric lapse using analytic or numerical differentiation.

We express a small radial perturbation in dimensionless form as
\begin{equation}
a(\tau)=a_0\bigl[1+\varepsilon y(\tau)\bigr],\qquad 0<|\varepsilon|\ll1,
\label{eq:perturbation_ansatz}
\end{equation}
where $y$ is dimensionless. Expanding the potential around $a_0$ gives
\begin{equation}
V(a)=\frac12V''_0a_0^2\varepsilon^2y^2+\mathcal{O}(\varepsilon^3),
\label{eq:potential_expansion}
\end{equation}
because the two equilibrium conditions remove the zeroth- and first-order terms. The linearized equation is
\begin{equation}
\ddot y+\frac12V''_0y=0.
\label{eq:linearized_stability}
\end{equation}
Thus $V''_0>0$ gives local oscillatory stability, $V''_0<0$ gives a growing linear mode, and $V''_0=0$ requires higher-order analysis. The last case is marginal at this order, while a negative $V''_0$ is a local statement and does not prove global collapse. Varying $a_0$ labels distinct static wormholes, whereas $\varepsilon y(\tau)$ describes the perturbation of one selected equilibrium.

\section{Equation-of-state stability sectors}\label{sec:eos_stability}

We consider three constitutive laws for the surface matter. Their original cosmological interpretations are therefore not transferred automatically to a two-dimensional throat. In each model, the equilibrium condition fixes one natural constitutive parameter, while the remaining parameters are held fixed during the perturbation of that equilibrium.

\subsection{Linear barotropic fluid}\label{subsec:barotropic}

The linear barotropic relation is the standard local constitutive approximation used in linearized thin-shell analyses, including the stability framework of Poisson and Visser \cite{PoissonVisser1995}:
\begin{equation}
p^{\rm(B)}=\omega\,\sigma^{\rm(B)}.
\label{eq:barotropic_eos}
\end{equation}
The parameter $\omega$ is dimensionless. A universal physical interval is not imposed by the barotropic relation itself; mathematically we take $\omega\in\mathbb{R}$, with admissible values constrained by the energy-condition and reality requirements of the throat. This use is phenomenological on the surface and does not assert a bulk fluid interpretation. The constitutive derivative is
\begin{equation}
\zeta^{\rm(B)}=\omega,\qquad p_a^{\rm(B)}=0.
\label{eq:barotropic_derivatives}
\end{equation}
The sign and magnitude of $\omega$ determine the local pressure response, with no explicit radius dependence. The constitutive law has no intrinsic density scale; that scale is supplied by the throat radius and the magnetic lapse.

The admissible interval is consequently selected by the calibrated throat rather than by an external cosmological prior. Admissible configurations have $a_0>0$, $\mathcal{F}_0>0$, real shell stresses, and a calibrated barotropic coefficient within the parameter interval under consideration. These conditions make the barotropic sector a controlled local benchmark for the magnetic geometry and keep its interpretation at the level of an effective surface law.

The limiting value $\omega=0$ recovers a pressureless surface response, whereas $\omega=1$ gives the stiff linear benchmark; these constitutive limits do not by themselves select an admissible static shell.

For a static configuration, the equilibrium stresses in \cref{eq:sigma_static,eq:p_static} fix the barotropic parameter as
\begin{equation}
\omega_0=\frac{p_0}{\sigma_0}=-\frac{2\mathcal{F}_0+a_0\mathcal{F}'_0}{4\mathcal{F}_0}.
\label{eq:barotropic_calibration}
\end{equation}
Substitution into \cref{eq:potential_second_static} gives the model-specific criterion
\begin{equation}
V_{0}^{\prime\prime\,\rm(B)}=\mathcal{F}''_0+\frac{\mathcal{F}'_0}{a_0}-\frac{(\mathcal{F}'_0)^2}{\mathcal{F}_0}.
\label{eq:barotropic_stability}
\end{equation}
The calibration in \cref{eq:barotropic_calibration} is held fixed when the selected static throat is perturbed. The sign of \cref{eq:barotropic_stability} is the barotropic benchmark for the two Chaplygin sectors.

\begin{figure*}[!htbp]
\centering
\includegraphics[width=\textwidth]{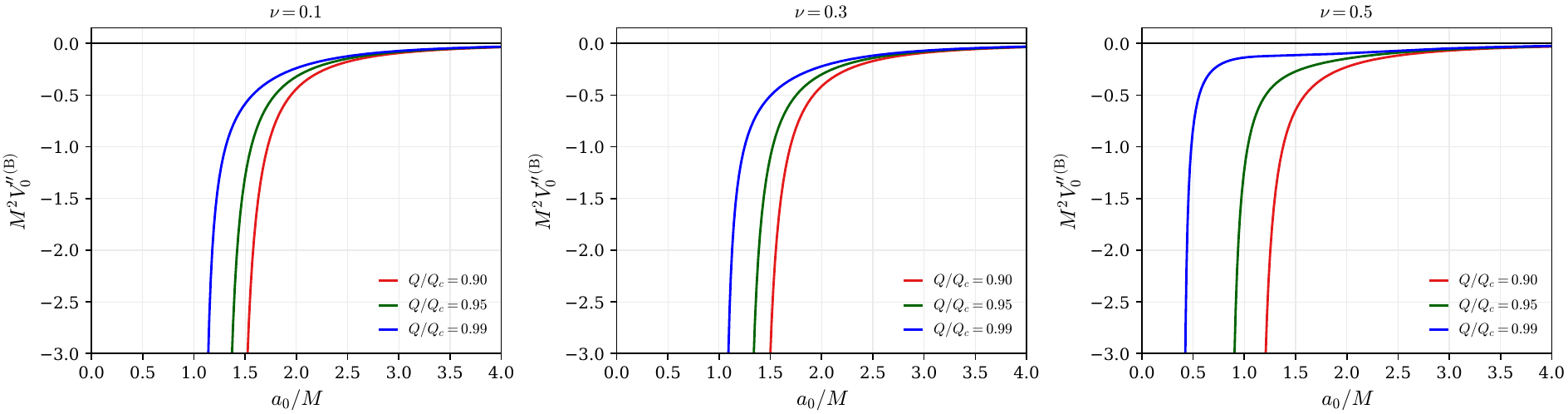}
\caption{Calibrated barotropic stability curvature $M^2 V_{0}^{\prime\prime\,\rm(B)}$ as a function of $a_0/M$ for the same fixed-mass parameter family as Fig.~\ref{fig:energy_subcritical_2x3}. The columns correspond to $\nu=0.1$, $0.3$, and $0.5$; red, green, and blue curves represent $Q/Q_c=0.90$, $0.95$, and $0.99$, respectively. The nonlinear magnetic scale obeys \cref{eq:mass_normalized_magnetic_constraint}. Only the exterior of the outer seed horizon within the applicable magnetic-field domain is retained. The black line marks zero curvature; the negative divergence near each simple horizon extends below the displayed ordinate range.}
\label{fig:barotropic_vpp_subcritical}
\end{figure*}

All nine curves in Fig.~\ref{fig:barotropic_vpp_subcritical} lie below zero throughout their displayed static domains. By the radial-mode criterion in Eq.~\eqref{eq:linearized_stability}, the calibrated barotropic layer therefore has a growing linear mode at every sampled throat in this parameter slice. In particular, the curvature remains negative on both sides of the tangential-NEC zero identified in Fig.~\ref{fig:energy_subcritical_2x3}; satisfaction of that tangential condition does not ensure radial stability. These statements concern local perturbations of distinct static junctions, each with its own coefficient from Eq.~\eqref{eq:barotropic_calibration}. They do not describe the evolution of one shell or imply inevitable global collapse.

Within the common exterior interval of each column, raising $Q/Q_c$ makes the curvature less negative and shifts the inner edge of the curve inward with the shrinking outer seed horizon. The first two columns have similar profiles away from those edges: at equal $Q/Q_c$, their plotted values differ by less than $6.2\times10^{-3}$ for $2.5\leq a_0/M\leq4$. Their common mass and nearby geometric charges give similar leading exterior terms in Eq.~\eqref{eq:rn_asymptotic}. The $\nu=0.5$ column shows a larger displacement of the near-horizon portions, especially for $Q/Q_c=0.99$; the apparent convergence at larger radii is not exact independence from $\nu$.

At a simple outer horizon, the final term of Eq.~\eqref{eq:barotropic_stability} drives $V_{0}^{\prime\prime\,\rm(B)}$ to negative infinity as the throat approaches the horizon from above. The vertical clipping in the figure resolves the finite exterior behavior and does not make the horizon an allowed timelike throat. The calibrated coefficient in Eq.~\eqref{eq:barotropic_calibration} also becomes unbounded in this limit, so the curves near different horizons should not be read as perturbations with one common barotropic coefficient. At large radius, Eqs.~\eqref{eq:rn_asymptotic} and \eqref{eq:barotropic_stability} give a leading negative inverse-cubic decay of the curvature. This analytic limit explains its approach toward zero from below; the plotted window alone does not establish the sign at every radius or for other parameter choices.

\subsection{Generalized cosmic Chaplygin gas}\label{subsec:gccg}

This cosmic extension of the inverse-density Chaplygin law adds a second nonlinear exponent \cite{GonzalezDiaz2003GCCG}. It has also been used as a surface closure for thin-shell wormholes \cite{SharifMumtaz2014CosmicChaplygin}. A regularized string-cloud construction provides another application \cite{ZhongEtAl2026RegularizedShell}; here the magnetic junction stresses determine the equilibrium parameters. To keep the surface dimensions explicit, fix a positive density scale $\sigma_\star=M^{-1}$ and write $\widehat\sigma=\sigma/\sigma_\star$. We adopt
\begin{equation}
p^{\rm(G)}=-\sigma_\star\widehat\sigma^{-\alpha_{\rm C}}
\left[E_{\rm G}+\left(\widehat\sigma^{1+\alpha_{\rm C}}-E_{\rm G}\right)^{-\varpi}\right],
\qquad E_{\rm G}=\frac{B_{\rm G}}{1+\varpi}-1.
\label{eq:gccg_eos}
\end{equation}
The Chaplygin exponent $\alpha_{\rm C}$, cosmic exponent $\varpi$, and coefficients $E_{\rm G}$ and $B_{\rm G}$ have no physical units; the density scale supplies the inverse-length dimension of the surface pressure. These symbols differ from the magnetic coupling $\gamma$, the barotropic coefficient $\omega$, and the shell area $\mathcal A_\Sigma$. The cosmological construction uses a positive volumetric density and a negative range $-\ell<\varpi<0$ with $\ell>0$; its direct $B_{\rm G}$ parametrization also excludes $\varpi=-1$. We use $0<\alpha_{\rm C}\leq1$ and retain only real static and perturbed surface states. Neither the cosmological density interpretation nor a fixed sign for $B_{\rm G}$ follows from this surface application.

The junction density is negative. Consequently $\alpha_{\rm C}$ must have an odd denominator when expressed as a reduced rational number; this specifies real powers of $\widehat\sigma<0$. We additionally require $Z_{\rm G}:=\widehat\sigma^{1+\alpha_{\rm C}}-E_{\rm G}>0$ in a neighborhood of each equilibrium. This positive-base condition excludes other algebraically real continuations for special integer values of $-\varpi$. The coefficient $B_{\rm G}$ is calibrated separately for each static throat and held fixed during its radial perturbation. At fixed constitutive parameters, direct differentiation gives
\begin{equation}
\zeta^{\rm(G)}=\varpi(1+\alpha_{\rm C})Z_{\rm G}^{-1-\varpi}
-\alpha_{\rm C}\frac{p^{\rm(G)}}{\sigma^{\rm(G)}},\qquad p_a^{\rm(G)}=0.
\label{eq:gccg_derivatives}
\end{equation}
The two terms in $\zeta^{\rm(G)}$ distinguish the nonlinear cosmic response from the inverse-density response. There is no independent radius dependence once the density scale and model parameters are fixed.

At fixed $B_{\rm G}$, the limit $\varpi\to0$ gives a generalized Chaplygin law on the selected real surface branch; the additional choice $\alpha_{\rm C}=1$ gives its ordinary inverse-density form. These are constitutive limits and do not establish stability without checking the corresponding calibrated equilibrium.

At a selected radius, let $Z_{{\rm G},0}=\widehat\sigma_0^{1+\alpha_{\rm C}}-E_{{\rm G},0}>0$. The static condition is equivalent to the positive-real root equation
\begin{equation}
Z_{{\rm G},0}-Z_{{\rm G},0}^{-\varpi}
=\widehat\sigma_0^{\alpha_{\rm C}}
\left(\widehat\sigma_0+\frac{p_0}{\sigma_\star}\right).
\label{eq:gccg_calibration}
\end{equation}
For every retained root, $E_{{\rm G},0}=\widehat\sigma_0^{1+\alpha_{\rm C}}-Z_{{\rm G},0}$ and $B_{{\rm G},0}=(1+\varpi)(E_{{\rm G},0}+1)$. Existence and uniqueness are not automatic; if several positive roots occur, a numerical family must state and follow its root-selection rule. Substitution into \cref{eq:potential_second_static} yields
\begin{equation}
V_{0}^{\prime\prime\,\rm(G)}=\mathcal F''_0-
\frac{(\mathcal F'_0)^2}{2\mathcal F_0}
+\left[1-2\alpha_{\rm C}\frac{p_0}{\sigma_0}
+2\varpi(1+\alpha_{\rm C})Z_{{\rm G},0}^{-1-\varpi}\right]
\left(\frac{\mathcal F'_0}{a_0}-\frac{2\mathcal F_0}{a_0^2}\right).
\label{eq:gccg_stability}
\end{equation}

\begin{figure}[!htbp]
\centering
\includegraphics[width=\textwidth]{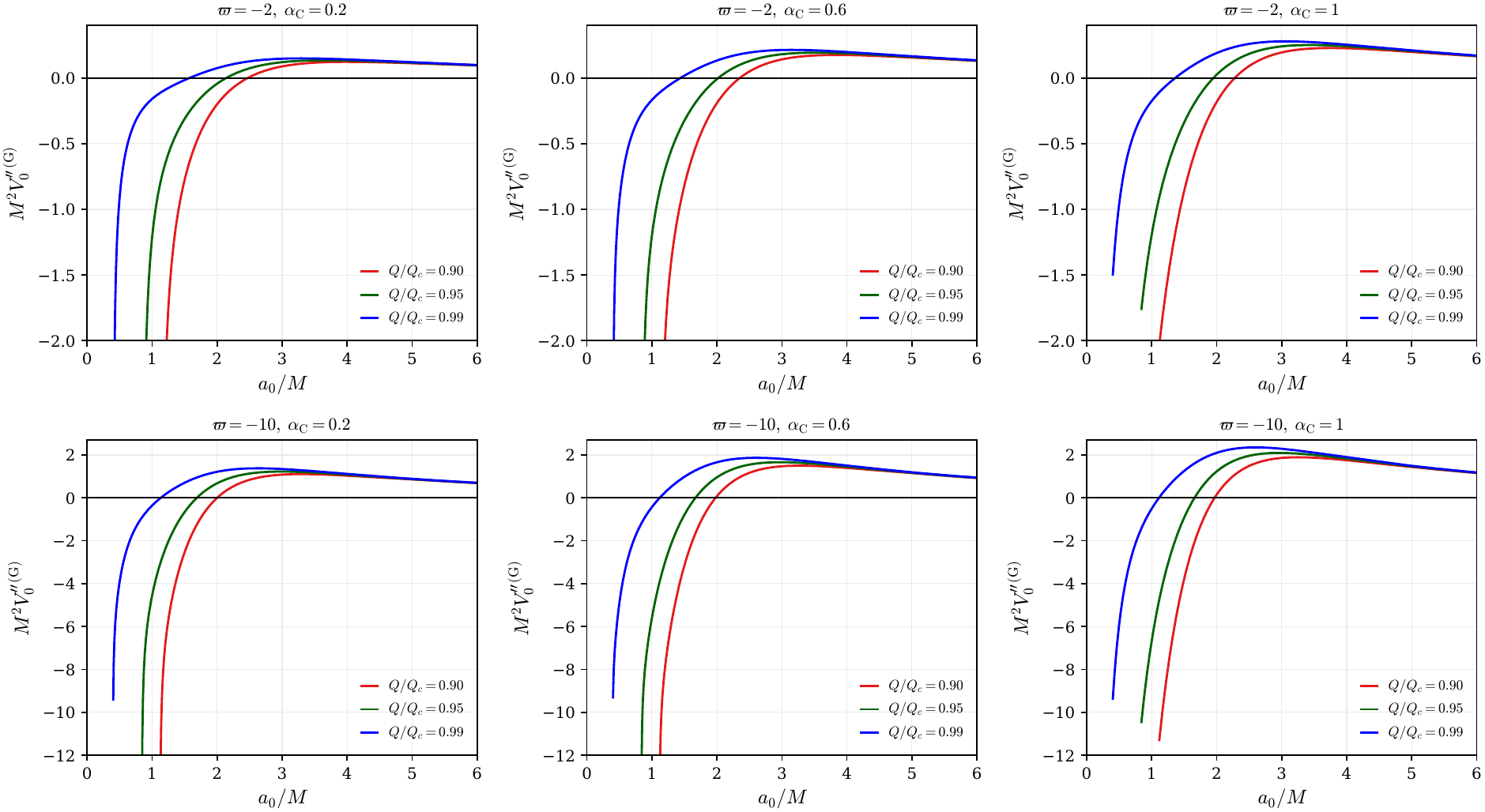}
\caption{GCCG radial stability curvature $M^2V_0^{\prime\prime\,\rm(G)}$ as a function of $a_0/M$ for $\nu=0.5$. Rows correspond to $\varpi=-2$ and $-10$; columns correspond to $\alpha_{\rm C}=0.2$, $0.6$, and $1$. Red, green, and blue curves represent $Q/Q_c=0.90$, $0.95$, and $0.99$, respectively, with the magnetic scale constrained by \cref{eq:mass_normalized_magnetic_constraint}. Each static throat is calibrated on the larger positive root of \cref{eq:gccg_calibration}, and its constitutive coefficients are fixed during the perturbation. Only the outer static exterior satisfying the magnetic-field bound is retained. A numerical sampling guard $(a_0-r_h)/M>0.01$ is applied. Horizontal black lines mark zero curvature; negative portions extending below the ordinate window are clipped.}
\label{fig:gccg_vpp_subcritical_nu05_2x3}
\end{figure}

Figure~\ref{fig:gccg_vpp_subcritical_nu05_2x3} contains a negative-curvature inner sector and a positive-curvature outer sector for every displayed GCCG family. This differs from the calibrated barotropic benchmark in Fig.~\ref{fig:barotropic_vpp_subcritical}. At fixed constitutive exponents, increasing $Q/Q_c$ moves the curvature zero inward. Increasing $\alpha_{\rm C}$ also moves this transition inward for the configurations considered, while the more negative cosmic exponent produces a larger positive maximum. The zeros span approximately $1.36\lesssim a_0/M\lesssim2.47$ in the upper row and $1.11\lesssim a_0/M\lesssim2.00$ in the lower row. These intervals locate the transitions among the selected families; they do not define a common stability boundary for all GCCG parameters.

The ordinate measures the local radial response defined by Eqs.~\eqref{eq:perturbation_ansatz} and \eqref{eq:linearized_stability}. The quantity $\varepsilon y(\tau)$ is a small fractional displacement of one equilibrium throat, expressed in shell proper time. A positive curvature gives a restoring frequency $\Omega_{\rm rad}=\sqrt{V''_0/2}$, whereas a negative curvature gives a growing-mode rate $\lambda_{\rm rad}=\sqrt{-V''_0/2}$; a zero requires higher-order analysis. Each horizontal position selects a different static throat and hence a different calibrated $B_{{\rm G},0}$. That coefficient, together with the other constitutive and bulk parameters, is fixed during the perturbation of the selected throat. The plotted sequence is therefore a family of local radial-mode tests, rather than a time evolution with a continually adjusted equation of state. Perturbations must remain within the same admissible exterior.

The distinct left boundaries originate in the charge-dependent outer seed horizons. In this fixed-mass family, $r_h/M\simeq1.107$, $0.831$, and $0.391$ for the three increasing charge ratios. The nonlinear magnetic scale changes with the charge according to Eq.~\eqref{eq:mass_normalized_magnetic_constraint}; it is not held independently fixed. A throat must lie strictly outside its corresponding horizon, and the sampling guard and vertical clipping can hide part of the negative branch. The stronger distinction between the first two columns and the $\alpha_{\rm C}=1$ column has an analytic origin. For a simple outer horizon with $\mathcal F'_h>0$, the static stresses imply $p_0/\sigma_0=-1/2-a_0\mathcal F'_0/(4\mathcal F_0)$. Substitution into Eq.~\eqref{eq:gccg_stability}, on the selected root and for $\varpi<-1$, gives
\begin{equation}
V_0^{\prime\prime\,\rm(G)}\sim
-\frac{(1-\alpha_{\rm C})\mathcal F'_h}{2(a_0-r_h)},
\qquad a_0\to r_h^+,\quad \alpha_{\rm C}\in\{0.2,0.6\}.
\label{eq:gccg_simple_horizon_limit}
\end{equation}
The cosmic response also grows in this limit, but more slowly than the displayed inverse-distance term. Thus the first two columns have a genuine negative divergence, rather than a finite discontinuity at the lower edge of the graph.

At $\alpha_{\rm C}=1$, the geometric and inverse-density contributions proportional to $(\mathcal F'_0)^2/\mathcal F_0$ cancel exactly. The calibration has a finite limiting root $Z_{{\rm G},h}>1$, and the curvature instead satisfies
\begin{equation}
\begin{aligned}
\lim_{a_0\to r_h^+}V_0^{\prime\prime\,\rm(G)}
&=\mathcal F''_h+
\left[1+4\varpi Z_{{\rm G},h}^{-1-\varpi}\right]
\frac{\mathcal F'_h}{r_h},\\
Z_{{\rm G},h}-Z_{{\rm G},h}^{-\varpi}
&=-\frac{\mathcal F'_h}{16\pi^2r_h\sigma_\star^2}.
\end{aligned}
\label{eq:gccg_unit_exponent_horizon_limit}
\end{equation}
This explains the finite beginnings of the third-column curves. The endpoint remains excluded from the timelike shell construction: the pressure still diverges, even though this particular curvature has a finite limit.

The positive outer sector follows from the competition between geometry and the constitutive response in Eq.~\eqref{eq:gccg_stability}. Where $a_0\mathcal F'_0<2\mathcal F_0$, the factor multiplying the cosmic response is negative. Since $\varpi<0$ and the calibrated base is positive, its contribution to the curvature is then positive. The change from $\varpi=-2$ to $-10$ strengthens this contribution in the selected families and raises their positive maxima. This dependence involves both the explicit cosmic exponent and the root fixed by Eq.~\eqref{eq:gccg_calibration}; it cannot be attributed to a charge-independent material constant.

At large radius, Eq.~\eqref{eq:rn_asymptotic} gives $p_0/\sigma_0\to-1/2$, while the larger calibration root approaches $Z_{{\rm G},0}\to1$. Consequently $\zeta^{\rm(G)}_0\to\varpi(1+\alpha_{\rm C})+\alpha_{\rm C}/2$, and
\begin{equation}
M^2V_0^{\prime\prime\,\rm(G)}\sim
\frac{2(1+\alpha_{\rm C})(-2\varpi-1)}{(a_0/M)^2},
\qquad a_0/M\to\infty.
\label{eq:gccg_large_radius_limit}
\end{equation}
The coefficient is positive for both cosmic exponents considered. This establishes a positive asymptotic tail on the chosen calibrated branch and explains why the charge curves approach one another within each panel: $Q$ and $\nu$ enter only at subleading orders. The common leading decay does not make the curves identical at finite radius. Their restoring frequency decreases as $a_0^{-1}$, so the approach toward zero corresponds to an increasingly slow local radial response, not to a finite-radius loss of stability inferred from the tail alone.

The bulk-exponent dependence complements this constitutive picture. At $\nu=0.1$ and $0.3$, the same charge ratios and GCCG exponents retain the negative inner sector, positive maximum, and decaying positive outer sector, but their curvature zeros occur at larger radii than for $\nu=0.5$. The two smaller bulk exponents give closely spaced transitions: for $\alpha_{\rm C}=1$, $\varpi=-10$, and $Q/Q_c=0.99$, the zero lies at $a_0/M\simeq2.019$ and $1.972$, respectively, compared with $1.112$ for $\nu=0.5$. These differences concern the mass-normalized magnetic family, in which $Q_c$ and the constrained nonlinear scale depend on $\nu$. They affect the finite-radius transition while leaving the leading asymptotic law in Eq.~\eqref{eq:gccg_large_radius_limit} unchanged.

\subsection{Modified cosmic Chaplygin gas}\label{subsec:mccg}

The modified cosmic Chaplygin gas (MCCG) adds a linear surface-pressure contribution to the same nonlinear family \cite{SharifMumtaz2014CosmicChaplygin,SadeghiFarahani2013MCCG}. We keep the density scale, exponents, and real-power convention of the GCCG sector, but use a distinct calibrated parameter $E_{\rm M}$:
\begin{equation}
p^{\rm(M)}=A_{\rm M}\sigma^{\rm(M)}-\sigma_\star\widehat\sigma^{-\alpha_{\rm C}}
\left[E_{\rm M}+\left(\widehat\sigma^{1+\alpha_{\rm C}}-E_{\rm M}\right)^{-\varpi}\right],
\qquad E_{\rm M}=\frac{B_{\rm M}}{1+\varpi}-1.
\label{eq:mccg_eos}
\end{equation}
Here $A_{\rm M}\geq0$ has no physical units, and $B_{\rm M}$ and $E_{\rm M}$ are dimensionless and need not be positive. The surface choice $A_{\rm M}\geq0$ selects a constitutive range rather than importing a cosmological prior. The local real domain requires $Z_{\rm M}:=\widehat\sigma^{1+\alpha_{\rm C}}-E_{\rm M}>0$. Setting $A_{\rm M}=0$ with $E_{\rm M}=E_{\rm G}$ recovers \cref{eq:gccg_eos}. Taking $\varpi\to0$ at fixed $B_{\rm M}$ gives the modified generalized Chaplygin relation; its inverse-density exponent becomes unity at $\alpha_{\rm C}=1$. A barotropic law follows from the combined $\varpi\to0$, $B_{\rm M}\to0$ limit, rather than from $B_{\rm M}=0$ alone at arbitrary $\varpi$.

With $A_{\rm M}$, $\alpha_{\rm C}$, $\varpi$, and $\sigma_\star$ fixed, the constitutive derivatives are
\begin{equation}
\zeta^{\rm(M)}=(1+\alpha_{\rm C})A_{\rm M}
+\varpi(1+\alpha_{\rm C})Z_{\rm M}^{-1-\varpi}
-\alpha_{\rm C}\frac{p^{\rm(M)}}{\sigma^{\rm(M)}},\qquad p_a^{\rm(M)}=0.
\label{eq:mccg_derivatives}
\end{equation}
The linear term affects both the explicit response and the equilibrium value of the nonlinear parameter. Defining $Z_{{\rm M},0}=\widehat\sigma_0^{1+\alpha_{\rm C}}-E_{{\rm M},0}>0$, the static pressure determines its admissible roots through
\begin{equation}
Z_{{\rm M},0}-Z_{{\rm M},0}^{-\varpi}
=\widehat\sigma_0^{\alpha_{\rm C}}
\left[(1-A_{\rm M})\widehat\sigma_0+\frac{p_0}{\sigma_\star}\right].
\label{eq:mccg_calibration}
\end{equation}
For a selected root, $E_{{\rm M},0}=\widehat\sigma_0^{1+\alpha_{\rm C}}-Z_{{\rm M},0}$ and $B_{{\rm M},0}=(1+\varpi)(E_{{\rm M},0}+1)$. The selected coefficient remains fixed during a perturbation; recalibration at another radius defines a different static shell. The local stability curvature is
\begin{equation}
V_{0}^{\prime\prime\,\rm(M)}=\mathcal F''_0-
\frac{(\mathcal F'_0)^2}{2\mathcal F_0}
+\left[1+2(1+\alpha_{\rm C})A_{\rm M}
-2\alpha_{\rm C}\frac{p_0}{\sigma_0}
+2\varpi(1+\alpha_{\rm C})Z_{{\rm M},0}^{-1-\varpi}\right]
\left(\frac{\mathcal F'_0}{a_0}-\frac{2\mathcal F_0}{a_0^2}\right).
\label{eq:mccg_stability}
\end{equation}
For $A_{\rm M}=0$ on the same calibrated real root, this reduces to \cref{eq:gccg_stability}. The magnetic lapse and its admissible exterior determine the sign of either curvature; the constitutive model alone does not fix a stability region.

\begin{figure}[!htbp]
\centering
\includegraphics[width=\textwidth]{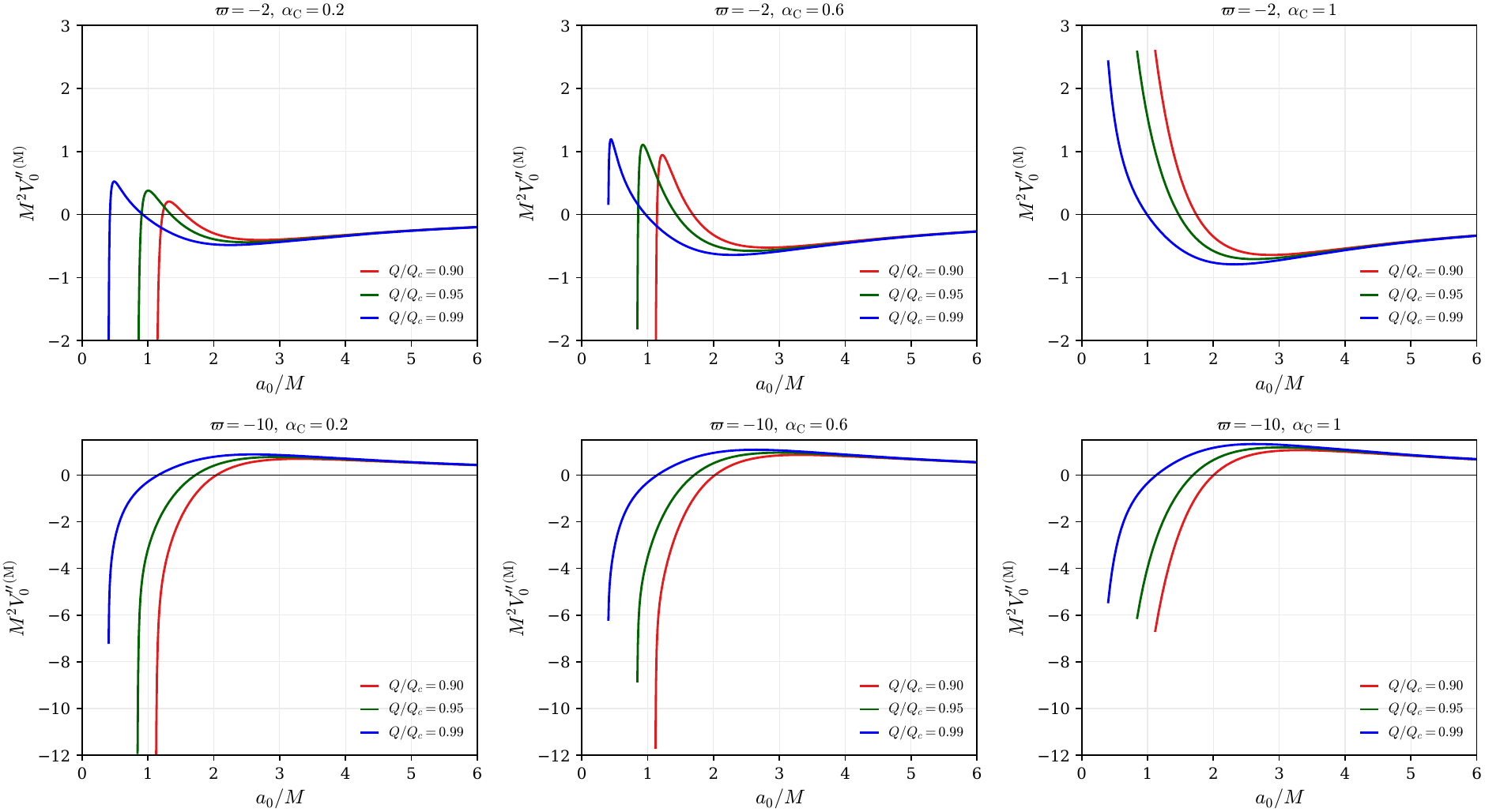}
\caption{MCCG radial stability curvature $M^2V_0^{\prime\prime\,\rm(M)}$ as a function of $a_0/M$ for $\nu=0.5$, $A_{\rm M}=4$, and $\sigma_\star=M^{-1}$. Rows correspond to $\varpi=-2$ and $-10$; columns correspond to $\alpha_{\rm C}=0.2$, $0.6$, and $1$. Red, green, and blue curves represent $Q/Q_c=0.90$, $0.95$, and $0.99$, respectively, with the magnetic scale constrained by \cref{eq:mass_normalized_magnetic_constraint}. Each static throat is calibrated on the larger positive root of \cref{eq:mccg_calibration}, and its constitutive coefficients are fixed during the perturbation. Only the outer static exterior satisfying the magnetic-field bound is retained, with a numerical sampling guard $(a_0-r_h)/M>0.01$. Horizontal black lines mark zero curvature. Ordinate limits are common within each row; negative portions below the window are clipped.}
\label{fig:mccg_vpp_nu05_AM4_2x3}
\end{figure}

Figure~\ref{fig:mccg_vpp_nu05_AM4_2x3} shows two distinct stability patterns. In the upper row, the fractional exponents produce finite positive-curvature intervals between an unstable inner sector and an unstable outer sector. For example, at $\alpha_{\rm C}=0.2$ these intervals are approximately $1.222<a_0/M<1.564$ for $Q/Q_c=0.90$ and $0.422<a_0/M<0.922$ for $Q/Q_c=0.99$. At $\alpha_{\rm C}=1$, the positive interval instead extends to the excluded horizon boundary, with one outer curvature zero. Increasing the charge ratio moves the upper-row boundaries inward for the selected families, while increasing $\alpha_{\rm C}$ moves their outer zero outward. The lower row retains the negative inner sector and positive outer sector found for the GCCG, with zeros spanning $1.135\lesssim a_0/M\lesssim2.042$. Here increasing either the charge ratio or $\alpha_{\rm C}$ moves the transition inward and raises the positive maximum. Thus the linear pressure contribution changes both the location and, in the upper row, the topology of the stable intervals.

The perturbation has the same meaning as in Eqs.~\eqref{eq:perturbation_ansatz} and \eqref{eq:linearized_stability}: $\varepsilon y(\tau)$ is the small fractional radial displacement of one equilibrium, and $\tau$ is shell proper time. Each ordinate is a local linear response with $A_{\rm M}$, $B_{{\rm M},0}$, the exponents, the density scale, and the seed parameters fixed. Recalibration along the horizontal axis selects different equilibria; it does not describe the evolution of one material law. Positive curvature gives an oscillatory restoring response and negative curvature gives an exponentially growing radial mode within this approximation. The curvature zeros require higher-order analysis and do not by themselves establish nonlinear stability.

The charge-dependent left boundaries are the same geometric boundaries as in Fig.~\ref{fig:gccg_vpp_subcritical_nu05_2x3}, since the surface law does not move the seed horizons. Their apparent locations also depend on the sampling guard and ordinate clipping. For the two fractional exponents, substituting the static stress ratio into Eq.~\eqref{eq:mccg_stability} leaves the singular contribution $-(1-\alpha_{\rm C})(\mathcal F'_0)^2/(2\mathcal F_0)$. At fixed finite $A_{\rm M}$, this dominates the other terms near a simple outer horizon, so the negative divergence in Eq.~\eqref{eq:gccg_simple_horizon_limit} also applies to the displayed MCCG fractional branches. The similarity of the first two columns therefore reflects a common singular mechanism, despite their different finite-radius maxima and stability boundaries.

At $\alpha_{\rm C}=1$, that singular term cancels. The $A_{\rm M}$ contribution to the calibration is proportional to $\widehat\sigma_0^2$ and vanishes at the horizon, giving the same limiting root as in Eq.~\eqref{eq:gccg_unit_exponent_horizon_limit}. Comparing the separately calibrated MCCG and GCCG at the same seed and exponents then yields
\begin{equation}
\lim_{a_0\to r_h^+}
\left[V_0^{\prime\prime\,\rm(M)}-V_0^{\prime\prime\,\rm(G)}\right]
=4A_{\rm M}\frac{\mathcal F'_h}{r_h},
\qquad \alpha_{\rm C}=1.
\label{eq:mccg_unit_exponent_horizon_shift}
\end{equation}
For $A_{\rm M}=4$, the resulting dimensionless MCCG limits lie between $2.58$ and $2.69$ in the upper row and between $-6.86$ and $-5.77$ in the lower row. These finite limits explain the distinct beginnings of the third-column curves. They do not make the endpoint a regular timelike throat: the pressure still diverges and $a_0=r_h$ remains excluded.

The effect of the linear coefficient includes the adjustment of the nonlinear calibration. At fixed seed, $a_0$, $\alpha_{\rm C}$, $\varpi$, and $\sigma_\star$, differentiating Eq.~\eqref{eq:mccg_calibration} along the selected simple root gives
\begin{equation}
\begin{aligned}
\partial_{A_{\rm M}}V_0^{\prime\prime\,\rm(M)}
&=2\left(\frac{\mathcal F'_0}{a_0}-\frac{2\mathcal F_0}{a_0^2}\right)
\partial_{A_{\rm M}}\zeta_0^{\rm(M)},\\
\partial_{A_{\rm M}}\zeta_0^{\rm(M)}
&=(1+\alpha_{\rm C})\left[
1-\frac{\varpi(1+\varpi)Z_{{\rm M},0}^{-2-\varpi}
\widehat\sigma_0^{1+\alpha_{\rm C}}}
{-\varpi Z_{{\rm M},0}^{-1-\varpi}-1}\right].
\end{aligned}
\label{eq:mccg_calibrated_linear_sensitivity}
\end{equation}
This derivative compares calibrated material laws and is distinct from the fixed-coefficient derivative during a radial perturbation. Its response factor is positive at the sampled equilibria for $A_{\rm M}=0.5,1,2,4,10$ and the charge ratios and exponents considered here. Increasing $A_{\rm M}$ therefore raises the curvature where $a_0\mathcal F'_0>2\mathcal F_0$ and lowers it where the opposite inequality holds. The geometric factor has the sign of the tangential NEC combination in Eq.~\eqref{eq:nec_combination}; where it vanishes, the curvature is independent of the constitutive response. This sign change explains how a stronger linear contribution can favor an inner stable interval while suppressing the outer one. It does not identify tangential NEC saturation with a stability boundary.

At large radius, $Z_{{\rm M},0}\to1$ and $p_0/\sigma_0\to-1/2$ imply $\zeta_0^{\rm(M)}\to(1+\alpha_{\rm C})(A_{\rm M}+\varpi)+\alpha_{\rm C}/2$. Hence
\begin{equation}
\begin{aligned}
M^2V_0^{\prime\prime\,\rm(M)}
&\sim\frac{2(1+\alpha_{\rm C})(-1-2A_{\rm M}-2\varpi)}{(a_0/M)^2},
\qquad a_0/M\to\infty,\\
A_{\rm M,crit}&=-\varpi-\tfrac12.
\end{aligned}
\label{eq:mccg_large_radius_limit}
\end{equation}
The leading coefficient changes sign at $A_{\rm M,crit}=1.5$ for $\varpi=-2$ and $9.5$ for $\varpi=-10$; at equality, subleading terms must determine the asymptotic classification. For the selected $A_{\rm M}=4$, the coefficients are $-10(1+\alpha_{\rm C})$ and $22(1+\alpha_{\rm C})$, respectively. This establishes the negative upper-row tail and positive lower-row tail beyond the plotted window. Charge and bulk-exponent dependence occurs only at subleading orders, explaining the convergence of the curves within each panel without implying equality at finite radius. The proper-time growth rate in the upper tail and restoring frequency in the lower tail both decrease as $a_0^{-1}$. Approaching zero from below therefore corresponds to a slower instability, rather than restoration of stability.

The dependence on $A_{\rm M}$ extends this distinction. At $A_{\rm M}=0.5$ and $1$, both cosmic exponents retain positive outer sectors on the same calibrated branch. At $A_{\rm M}=2$, the $\varpi=-2$ tail is already negative; for $\alpha_{\rm C}=0.2$ none of the three selected charge ratios has a positive interval over $r_h+0.01M<a_0\leq100M$, whereas finite positive intervals survive for some larger exponents. For $\varpi=-10$, the positive maximum at $\alpha_{\rm C}=1$ and $Q/Q_c=0.99$ decreases from approximately $2.23$ at $A_{\rm M}=0.5$ to $2.10$, $1.84$, and $1.33$ at $A_{\rm M}=1,2,4$, respectively. At $A_{\rm M}=10$, this row also has a negative asymptotic tail, although finite positive-curvature intervals can persist. These results show that increasing the linear pressure coefficient does not uniformly improve radial stability; its effect depends on the geometric response factor and on the balance with the calibrated cosmic term.

\section{Thermodynamics of the magnetic throat}\label{sec:thermodynamics}

\subsection{Shell energy and the first-law balance}\label{subsec:thermal_balance}

The surface stresses determine the mechanical energy of the layer, whereas a thermodynamic interpretation additionally requires a state space and a temperature prescription. We adopt the shell first-law framework developed by Lobo and Rodrigues \cite{LoboRodriguesThermodynamics2026}, keeping the magnetic geometry, the surface constitutive parameters, and any conserved shell particle number fixed during a physical radial evolution.

The area and proper internal energy of the shell are
\begin{equation}
\mathcal{A}_{\Sigma}=4\pi a^2,\qquad E_{\Sigma}=\mathcal{A}_{\Sigma}\sigma.
\label{eq:shell_thermal_variables}
\end{equation}
Here $E_{\Sigma}$ denotes the energy measured in the shell rest frame; it is distinct from the positive magnetic ADM mass at either asymptotic end. Substitution of the dynamical junction density gives
\begin{equation}
E_{\Sigma}=-2a\sqrt{\mathcal{F}(a)+\dot a^2}.
\label{eq:shell_thermal_energy}
\end{equation}
The wormhole orientation fixes the negative sign of the shell internal energy. Replacing $E_{\Sigma}$ by its magnitude would change the pressure-work term and the adopted first law.

With a possible normal energy flux, the intrinsic balance takes the form
\begin{equation}
\frac{\dd E_{\Sigma}}{\dd\tau}+p\frac{\dd\mathcal{A}_{\Sigma}}{\dd\tau}=\mathcal{Q}_{\Sigma}.
\label{eq:shell_energy_balance}
\end{equation}
The quantity $\mathcal{Q}_{\Sigma}$ is the net bulk energy delivered to the layer per unit proper time. The magnetic branch satisfies $T^t{}_{t}=T^r{}_{r}$, equivalently $\rho+p_r=0$, so its normal contraction with the radial shell velocity vanishes on each side. Together with the electromagnetic matching described in \cref{sec:construction}, this gives $\mathcal{Q}_{\Sigma}=0$ even for radial motion. This cancellation follows from the magnetic stress tensor and does not require a vacuum exterior.

For the magnetic junction, the tangential monopole two-form $F=q\sin\theta\,\dd\theta\wedge\dd\varphi$ is identified continuously across the throat and its contraction with the radial normal vanishes. The corresponding nonlinear electromagnetic displacement has no normal component, so no surface electric current is introduced by this matching. Charges measured with outward orientations at the two asymptotic ends have opposite signs, while the geometry depends on $q^2$. These conditions preserve the magnetic flux and supply the field matching used in the zero-flux balance.

For a reversible local equilibrium description, the assumed first law is
\begin{equation}
T_{\Sigma}\dd S_{\Sigma}=\dd E_{\Sigma}+p\dd\mathcal{A}_{\Sigma}.
\label{eq:shell_first_law}
\end{equation}
Chemical and charge-work terms are absent in this variation because the corresponding conserved quantities are held fixed. The law is an additional thermodynamic assumption; the junction equations alone do not determine $S_{\Sigma}$ or identify the thermodynamic temperature $T_{\Sigma}$.

Along the transparent radial evolution, \cref{eq:shell_energy_balance,eq:shell_first_law} imply
\begin{equation}
T_{\Sigma}\frac{\dd S_{\Sigma}}{\dd\tau}=0.
\label{eq:shell_adiabaticity}
\end{equation}
For $T_{\Sigma}>0$, the motion is isentropic. This establishes conservation of entropy, rather than its absolute value. At zero temperature the product vanishes without fixing the entropy derivative, and a separate limiting prescription is required.

The same cancellation can be checked directly on the static geometric locus. Differentiating the static specialization of \cref{eq:shell_thermal_energy} at fixed seed parameters and combining it with \cref{eq:p_static} gives
\begin{equation}
\mathcal{W}_{a}:=\frac{\dd E_{\Sigma,0}}{\dd a_0}+p_0\frac{\dd\mathcal{A}_{\Sigma,0}}{\dd a_0}=0.
\label{eq:static_thermal_work}
\end{equation}
The result relies on the areal angular function $r^2$. It differs from nonareal geometries, for which an additional geometric work coefficient can survive. The static junction locus comprises different equilibria; when an EOS coefficient is recalibrated at each radius, this locus is not automatically a trajectory of one fixed thermodynamic substance. Equation \eqref{eq:static_thermal_work} therefore does not assign a common entropy to all independently calibrated EOS models.

\subsection{Thermodynamic completions of the three surface equations of state}\label{subsec:thermal_eos}

A surface EOS fixes a pressure response but leaves a functional freedom in the entropy dependence of the internal energy. This can be made explicit without introducing another matter model. Write $x=-\sigma>0$ and use $K_{\rm B}(S_\Sigma)$ for the barotropic integration function. Any fixed conserved particle number or magnetic-background labels are suppressed in the entropy functions below.

On an isentrope, \cref{eq:shell_first_law} becomes
\begin{equation}
\mathcal{A}_{\Sigma}\left(\frac{\partial x}{\partial\mathcal{A}_{\Sigma}}\right)_{S_{\Sigma}}=p-x.
\label{eq:eos_isentrope}
\end{equation}
This relation preserves the negative-density branch used in the stability calculation. Integrating the barotropic law at fixed $\omega$ yields the fundamental relation
\begin{equation}
E_{\Sigma}^{\rm(B)}(S_{\Sigma},\mathcal{A}_{\Sigma})
=-K_{\rm B}(S_{\Sigma})\mathcal{A}_{\Sigma}^{-\omega}.
\label{eq:barotropic_fundamental}
\end{equation}
Positive $K_{\rm B}$ gives negative energy and real density for every positive area. The function has the dimensions required by the area factor and is calibrated at the selected equilibrium; its entropy derivatives are independent thermodynamic information.

For $j\in\{{\rm G},{\rm M}\}$, let $p^{(j)}(\sigma)$ denote the respective cosmic Chaplygin relation in \cref{eq:gccg_eos,eq:mccg_eos} with all constitutive coefficients held fixed. On any connected real branch where $\sigma+p^{(j)}(\sigma)\ne0$, integration of \cref{eq:eos_isentrope} gives an implicit fundamental relation:
\begin{equation}
\ln\frac{\mathcal A_\Sigma}{\mathcal A_\star}
=H_j(S_\Sigma)-\int_{\sigma_{\rm ref}}^{\sigma}
\frac{\dd s}{s+p^{(j)}(s)},\qquad
E_\Sigma^{(j)}=\mathcal A_\Sigma\sigma.
\label{eq:cosmic_chaplygin_fundamental}
\end{equation}
Here $\mathcal A_\star>0$ is a fixed reference area, $\sigma_{\rm ref}<0$ lies on the same admissible branch, and $H_j$ is an undetermined dimensionless function of entropy. The formula is local: it cannot be continued across a zero of $\sigma+p^{(j)}$, a zero of $Z_j$, or a loss of the selected real power. At $\sigma+p^{(j)}=0$, a constant-density isentrope requires separate treatment. Thus the two cosmic closures determine pressure and an isentropic relation but do not determine a unique entropy dependence or a global thermal state space.

The conjugate variables are fixed by
\begin{equation}
T_{\Sigma}=\left(\frac{\partial E_{\Sigma}}{\partial S_{\Sigma}}\right)_{\mathcal{A}_{\Sigma}},\qquad
p=-\left(\frac{\partial E_{\Sigma}}{\partial\mathcal{A}_{\Sigma}}\right)_{S_{\Sigma}}.
\label{eq:thermal_conjugates}
\end{equation}
The second identity recovers the barotropic and both cosmic Chaplygin surface laws on their stated local branches. For the barotropic completion, $K'_{\rm B}<0$ gives positive temperature. For either cosmic law, differentiation of \cref{eq:cosmic_chaplygin_fundamental} at fixed area gives $T_\Sigma=\mathcal A_\Sigma[\sigma+p^{(j)}(\sigma)]H'_j(S_\Sigma)$; its sign depends on both factors. The equilibrium density fixes one value of the integration function, while a temperature prescription constrains a first derivative. Neither fixes the full entropy dependence or the heat capacity.

For example, a heat capacity at fixed area would be
\begin{equation}
C_{\mathcal{A}}=\frac{T_{\Sigma}}{(\partial T_{\Sigma}/\partial S_{\Sigma})_{\mathcal{A}_{\Sigma}}}.
\label{eq:shell_heat_capacity}
\end{equation}
It is defined only when the denominator is nonzero and depends on $K_{\rm B}$ or the relevant $H_j$. The zero work in \cref{eq:static_thermal_work} therefore cannot be interpreted as a zero heat capacity. A conventional local thermal-stability test in the entropy representation additionally requires
\begin{equation}
S_{EE}\leq0,\qquad S_{\mathcal{A}\mathcal{A}}\leq0,
\label{eq:entropy_concavity_diagonal}
\end{equation}
and
\begin{equation}
S_{EE}S_{\mathcal{A}\mathcal{A}}-S_{E\mathcal{A}}^2\geq0.
\label{eq:entropy_concavity_determinant}
\end{equation}
These derivatives are taken at fixed conserved quantities in an admissible, invertible state-space patch. They are conditional criteria for the adopted thermodynamic description. They remain distinct from the dynamical condition in \cref{eq:linearized_stability}, which tests a radial mode at fixed EOS parameters. Without a microscopic entropy prescription, radial stability alone does not determine a global thermal-stability region.

\subsection{Acceleration temperature and equilibrium redshift}\label{subsec:acceleration_temperature}

Following the local acceleration construction of Lobo and Rodrigues \cite{LoboRodriguesThermodynamics2026,LoboRodriguesUnruh2026}, the invariant acceleration magnitude of the moving shell is
\begin{equation}
\alpha_{\Sigma}=\frac{|\ddot a+\mathcal{F}'(a)/2|}{\sqrt{\mathcal{F}(a)+\dot a^2}}.
\label{eq:shell_proper_acceleration}
\end{equation}
The two identical exteriors give the same magnitude, so averaging their contributions introduces no additional factor. The corresponding local acceleration-temperature scale is
\begin{equation}
T_{\rm acc}=\frac{\hbar\alpha_{\Sigma}}{2\pi}.
\label{eq:shell_acceleration_temperature}
\end{equation}
Identifying $T_{\Sigma}$ in \cref{eq:shell_first_law} with this scale is a thermodynamic prescription. An actual thermal detector response further depends on the field state and trajectory; an acceleration formula by itself does not prove a Planck spectrum in this curved background.

At the static magnetic throat, the prescription becomes
\begin{equation}
T_{{\rm acc},0}=\frac{\hbar|\mathcal{F}'_0|}{4\pi\sqrt{\mathcal{F}_0}}.
\label{eq:static_acceleration_temperature}
\end{equation}
This expression uses the exact Kruglov lapse and the same positive-lapse and magnetic-field domain as the junction stresses. At a lapse extremum it vanishes; dividing a first-law balance by it would be undefined there. Its sign is nonnegative regardless of the sign of the radial acceleration.

For a genuine static thermal equilibrium with a reservoir, the Tolman relation is
\begin{equation}
T_{\rm eq}(a_0)\sqrt{\mathcal{F}_0}=T_{\infty}.
\label{eq:tolman_shell_equilibrium}
\end{equation}
The constant $T_{\infty}$ refers to a single stationary bath normalized by the asymptotic Killing time. Identifying $T_{\rm eq}=T_{{\rm acc},0}$ at one radius requires $T_{\infty}=\hbar|\mathcal{F}'_0|/(4\pi)$. Since this right-hand side generally changes with $a_0$, a radius scan of the acceleration temperature describes separately assigned local states and does not generally describe one bath in equilibrium with all shells. In particular, local acceleration temperature and a Tolman-redshifted Hawking temperature are not equal at arbitrary exterior radii.

Suppose the seed has a simple outer Killing horizon $r_h$ with $\mathcal{F}'_h>0$, and approach it from the retained exterior. For $\delta=a_0-r_h>0$, the leading static behavior is
\begin{equation}
T_{{\rm acc},0}\sim\frac{\hbar\sqrt{\mathcal{F}'_h}}{4\pi\sqrt{\delta}}.
\label{eq:simple_horizon_acceleration_limit}
\end{equation}
The proper distance from the seed horizon is $\ell_h\sim2\sqrt{\delta/\mathcal{F}'_h}$, so the same limit is $T_{{\rm acc},0}\sim\hbar/(2\pi\ell_h)$. It agrees at leading order with the blueshift of the seed Hawking temperature. This asymptotic statement requires a simple root and does not turn the excised horizon into a horizon of the static wormhole.

If an admissible extremal sector instead has $\mathcal{F}(r)=C_h(r-r_h)^2+o((r-r_h)^2)$ with $C_h>0$, then
\begin{equation}
\lim_{a_0\to r_h^+}T_{{\rm acc},0}=\frac{\hbar\sqrt{C_h}}{2\pi}.
\label{eq:double_horizon_acceleration_limit}
\end{equation}
The seed Hawking temperature is zero in this double-root limit, while the local static acceleration scale need not vanish. The extremal and simple-root limits therefore require separate treatment. The existence of such a positive-radius double root must be established in the chosen magnetic parameter sector; a horizon disappearing at $r=0$ is not this limit.

\subsection{Temperature response to a radial mode}\label{subsec:thermal_radial_response}

The acceleration prescription also gives a local temperature response to the radial perturbation already defined in \cref{eq:perturbation_ansatz}. Consider $\mathcal{F}'_0\ne0$ and a small displacement $\delta a=a_0\varepsilon y$, with the seed parameters held fixed and the acceleration numerator retaining its equilibrium sign. Expanding \cref{eq:shell_proper_acceleration,eq:shell_acceleration_temperature} to first order gives
\begin{equation}
\frac{\delta T_{\rm acc}}{T_{{\rm acc},0}}
=\frac{2\delta\ddot a}{\mathcal{F}'_0}
+\left(\frac{\mathcal{F}''_0}{\mathcal{F}'_0}-\frac{\mathcal{F}'_0}{2\mathcal{F}_0}\right)\delta a.
\label{eq:temperature_radial_kinematic_response}
\end{equation}
The velocity-squared term enters only at second order. The first term is essential for a moving throat and would be lost by differentiating only the static temperature curve.

Using the mechanical linearized equation yields
\begin{equation}
\frac{\delta T_{\rm acc}}{T_{{\rm acc},0}}
=\left[\frac{\mathcal{F}''_0-V''_0}{\mathcal{F}'_0}-\frac{\mathcal{F}'_0}{2\mathcal{F}_0}\right]\delta a.
\label{eq:temperature_radial_mode_response}
\end{equation}
Each of the three EOS enters through its already calibrated $V''_0$, with its constitutive coefficients fixed during the perturbation. In a stable sector, the induced temperature variation oscillates with the radial-mode frequency unless its linear coefficient vanishes. This is a kinematic response of the adopted acceleration-temperature scale; it does not replace the entropy-concavity criteria or determine a thermal relaxation rate. At $\mathcal{F}'_0=0$, the absolute value makes this fractional expansion unavailable, and the unexpanded acceleration formula must be used. Identifying this entire dynamical response with a state-function temperature of shell matter would require an additional nonequilibrium constitutive prescription.

\subsection{Seed-parameter variations and horizon comparisons}\label{subsec:thermal_parameter_variations}

Thermodynamic parameter variations refer to the magnetic family defined in \cref{sec:seed}.

Let $\lambda$ denote one independent seed label under an explicitly chosen variation prescription. At fixed $a_0$, the static energy gives
\begin{equation}
\dd E_{\Sigma,0}+p_0\dd\mathcal{A}_{\Sigma,0}
=-\frac{a_0}{\sqrt{\mathcal{F}_0}}
\left(\frac{\partial\mathcal{F}_0}{\partial\lambda}\right)_{a_0}\dd\lambda.
\label{eq:seed_parameter_work}
\end{equation}
The derivative incorporates \cref{eq:magnetic_self_mass} or its fixed-mass form \cref{eq:mass_normalized_magnetic_constraint}, according to the chosen independent variation. This is a mechanical energy change under alteration of the exterior geometry. Assigning it entirely to $T_{\Sigma}\dd S_{\Sigma}$ requires an ensemble prescription and an accounting of the external work; otherwise generalized conjugate-work terms must be included. The magnetic flux of the background is also distinct from a material charge residing on the shell. No Smarr relation is inferred from this coefficient alone.

For comparison only, a seed horizon in four-dimensional Einstein gravity with the minimally coupled nonlinear electromagnetic Lagrangian has
\begin{equation}
T_h=\frac{\hbar|\mathcal{F}'_h|}{4\pi},\qquad S_h=\frac{\pi r_h^2}{\hbar}.
\label{eq:seed_horizon_thermodynamics}
\end{equation}
The entropy is the Einstein area entropy: the matter Lagrangian contains no curvature coupling that would modify its Wald factor. These quantities refer to the seed black hole and are defined only for existing, correctly classified horizons. The static wormhole retains $r\geq a_0>r_h$ and excises this horizon, so $S_h$ cannot be assigned to the throat. A meaningful shell--horizon entropy comparison would additionally require fixing the undetermined shell entropy and declaring the same independent parameter variation on both objects.

\subsection{Particle creation and the stationary limit}\label{subsec:thermal_peeling}

Particle creation depends on the mapping between incoming and outgoing null rays, in addition to a local acceleration scale \cite{LoboRodriguesUnruh2026}. Denote an incoming affine null coordinate by $U$ and an outgoing asymptotic retarded coordinate by $u$, with $U=P(u)$. The peeling scale is
\begin{equation}
\kappa_{\rm peel}(u)=-\frac{\dd}{\dd u}\ln P'(u).
\label{eq:peeling_scale}
\end{equation}
It probes the evolution of neighboring rays between asymptotic regions. It is not fixed by the junction pressure or by \cref{eq:shell_proper_acceleration}. With a suitable vacuum prescription and approximately exponential peeling, an adiabatic particle-creation temperature can be associated with
\begin{equation}
T_{\rm peel}=\frac{\hbar|\kappa_{\rm peel}|}{2\pi},\qquad
\left|\frac{\dd\kappa_{\rm peel}}{\dd u}\right|\ll\kappa_{\rm peel}^2.
\label{eq:peeling_temperature_condition}
\end{equation}
The temperature here is tied to the normalization of the asymptotic null time. A comparison with the local shell temperature must transport both temperatures to the same observer and specify the state and dynamical ray map; their equality is an additional condition.

For the stationary, transparent, symmetric construction with aligned asymptotic clocks, radial rays have a time-independent transit delay. Their map is affine, $P(u)=u+\mathrm{constant}$, giving $\kappa_{\rm peel}=0$. In a stationary vacuum this supplies no Hawking-like particle-creation flux, even when \cref{eq:static_acceleration_temperature} is positive. A nonstationary throat or a different quantum state would require a separate semiclassical calculation and, if appreciable radiation is produced, a corresponding change to the zero-flux and test-background assumptions. The optical disk emission studied next is a prescribed classical source and is independent of this particle-creation question.

\section{Optical appearance of bilateral disk sources}\label{sec:optical_appearance}

\subsection{Probe radiation, source geometry, and throat matching}\label{subsec:optical_setup}

We study the optical appearance of a stationary throat at a selected admissible radius $a_0$, using the exact magnetic background in \cref{eq:seed_metric,eq:seed_lapse}. The emitting matter consists of geometrically thin, transparent equatorial test disks in both exteriors. The observer resides in $\mathcal{M}_1$, and $\mathcal{M}_2$ is the opposite universe. The disk planes are aligned through the throat's angular identification; both disks use the same model-C radial profile, with explicit intensity normalizations, outer supports and rotation senses. Reflection symmetry of the background does not require equal illumination. The observer is static in $\mathcal M_1$ at $r_o=10^5M$, with $\mathcal F_o=\mathcal F(r_o)>0$. We compare static and moving emitters face-on, and retain the moving prescription for inclined viewing. The asymptotic-observer limit is obtained as $r_o\to\infty$ and $\mathcal F_o\to1$. Disk stresses, radiation, absorption, and scattering do not backreact on the metric or on the shell. Transparency is assumed for the probe radiation at the throat, independently of the mechanical EOS of the supporting layer.

The rays in this section are minimally coupled Maxwell test radiation, so their characteristics are null geodesics of the spacetime metric. This defines the optical model in addition to the nonlinear electromagnetic field supporting the background. We do not identify these rays with arbitrary polarization-dependent disturbances of the background nonlinear theory. Such an identification would require deriving that theory's effective characteristic cones and their junction conditions. In particular, independence of the magnetic metric from $\gamma$ does not by itself establish independence of a nonlinear electromagnetic propagation law from $\gamma$.

For the static matching, the proper time on the shell satisfies $\dd\tau=\sqrt{\mathcal{F}(a_0)}\dd t_i$ on either side. Let $E_i=-k_{t_i}$ be the photon Killing energy and $L$ its conserved total angular momentum. Continuity of the locally measured frequency and of tangential momentum gives
\begin{equation}
\frac{E_1}{\sqrt{\mathcal{F}(a_0)}}=\frac{E_2}{\sqrt{\mathcal{F}(a_0)}},\qquad L_1=L_2.
\label{eq:photon_throat_matching}
\end{equation}
The equality follows from the identical static exteriors and the assumed transparent junction. With aligned Killing-time normalizations, $E_1=E_2$ and the impact parameter $b=L/E\geq0$ is unchanged on crossing the throat. The radial component is continuous when expressed in an oriented proper-distance coordinate; in the separate areal coordinates its sign reverses because both coordinates increase away from the shell. No extra gravitational frequency jump is introduced by the symmetric matching.

The optical calculation is conditioned on a throat that can be treated as static during photon propagation. A radius satisfying $V''_0>0$ for an admissible EOS supplies the mechanical sector relevant to a persistent image. An unstable static junction can be used as a geometric diagnostic, but its image is not evidence of a long-lived configuration. The disk velocity and its emissivity are test-source assumptions and are not inferred from the barotropic or Chaplygin EOS of the shell.

\subsection{Null orbits and the imaging sector}\label{subsec:optical_null_orbits}

Spherical symmetry confines each ray to an orbital plane. If $\psi$ is the angle in that plane and $\lambda_{\rm aff}$ is an affine parameter, the radial and angular first integrals are
\begin{equation}
\left(\frac{\dd r}{\dd\lambda_{\rm aff}}\right)^2=E^2\left[1-b^2\mathcal{U}(r)\right],\qquad
\mathcal{U}(r)=\frac{\mathcal{F}(r)}{r^2},
\label{eq:optical_null_potential}
\end{equation}
and
\begin{equation}
\left|\frac{\dd\psi}{\dd r}\right|=\frac{b}{r^2\sqrt{1-b^2\mathcal{U}(r)}}.
\label{eq:optical_angle_integral}
\end{equation}
The square root specifies the accessible ray domain. Angular advance is accumulated monotonically along the backward-traced path, including both radial branches when an exterior turning point exists.

A circular null orbit away from the shell obeys
\begin{equation}
r\mathcal{F}'-2\mathcal{F}=0.
\label{eq:photon_sphere_equation}
\end{equation}
Each admissible root must be classified using $\mathcal{U}''$: negative curvature gives an unstable orbit and positive curvature a stable one. We denote the selected exterior unstable orbit by $r_{\rm ph}$. Its critical impact parameter is
\begin{equation}
b_c=\frac{r_{\rm ph}}{\sqrt{\mathcal{F}(r_{\rm ph})}}.
\label{eq:optical_critical_impact}
\end{equation}
The relevant orbit is determined by the exterior potential barrier accessible from infinity. A throat may also create an extremum or cusp of the global potential when written across the two joined regions. These features must be included when determining which ray branch reaches the shell.

The standard imaging sector used below requires
\begin{equation}
a_0<r_{\rm ph}<r_{\rm I},\qquad
\max_{a_0\leq r\leq r_{\rm ph}}\mathcal{U}(r)\leq\mathcal{U}(r_{\rm ph}),
\label{eq:optical_barrier_sector}
\end{equation}
where $r_{\rm I}$ is the physical outer innermost stable circular orbit (ISCO) introduced below. The retained interval must also lie in the outer static component and respect the magnetic-field bound. The second condition excludes an additional inner barrier higher than the selected photon barrier. If a parameter choice violates these conditions, its radial branches must be reconsidered before applying this sector's source and transfer construction.

For $b>b_c$, a backward ray turns in $\mathcal{M}_1$ and can intersect the same disk on its inward and outward legs. For $0\leq b<b_c$, it reaches the throat and continues into the emitting exterior $\mathcal M_2$. The identical barrier in $\mathcal{M}_2$ cannot reflect that subcritical ray. In an asymmetric construction, a ray can cross the throat, turn on the second side, and return to the emitting side in an interval between two distinct mapped critical impacts. Here both critical impacts are exactly $b_c$, so that returned interval has zero width. This does not remove photons emitted in $\mathcal M_2$ and transmitted to the observer: a future photon from that disk travels inward on side 2, crosses the throat once, and travels outward on side 1. The critical orbit itself is a limiting ray of infinite angular advance, not an ordinary finite intersection sequence. Consequently $b_c$ identifies the geometric critical curve, while the radius of a dark disk-image region also depends on source support and frequency shifts.

\subsection{Static emitters, circular motion, and the inward ISCO separatrix}\label{subsec:optical_gas_motion}

The face-on analysis first uses static emitters as a control for the frequency effects of motion. In either exterior their four-velocity is
\begin{equation}
u_{{\rm stat},i}^{\mu}=\mathcal F(r)^{-1/2}(1,0,0,0).
\label{eq:static_emitter_velocity}
\end{equation}
This velocity is future-directed and has unit timelike norm wherever $\mathcal F>0$. Maintaining it generally requires acceleration; it is a prescribed stationary source, not a geodesic accretion flow. The model-C profile retains its geometric ISCO scale in this control, without assigning orbital motion to the static matter.

The moving prescription is then used for both face-on and inclined observations. Each disk has its own test flow. At large radii the gas follows circular timelike geodesics with a specified rotation sense; below the physical ISCO it follows an inward pressure-free separatrix with conserved ISCO energy and angular momentum. This is a specified test-flow approximation, not a hydrodynamic accretion solution.

Introduce the circular normalization function
\begin{equation}
D(r)=\mathcal{F}(r)-\frac{r\mathcal{F}'(r)}2.
\label{eq:circular_normalization}
\end{equation}
Circular motion requires $\mathcal{F}>0$, $\mathcal{F}'>0$, and $D>0$. In the $(t,r,\theta,\varphi)$ coordinates of the emitting exterior, its velocity is
\begin{equation}
u_{{\rm circ},i}^{\mu}=D(r)^{-1/2}\bigl(1,0,0,s_i\Omega(r)\bigr),\qquad
\Omega(r)=\sqrt{\frac{\mathcal{F}'(r)}{2r}}.
\label{eq:circular_emitter_velocity}
\end{equation}
The positive square root defines the magnitude; $s_i=\pm1$ specifies the rotation sense in exterior $i$. The equal-illumination control uses $s_1=s_2=+1$. This four-velocity has unit timelike norm; $\Omega$ is a coordinate angular velocity and has inverse-length dimensions.

The marginal-stability equation is
\begin{equation}
r\mathcal{F}\mathcal{F}''+3\mathcal{F}\mathcal{F}'-2r(\mathcal{F}')^2=0.
\label{eq:optical_isco_equation}
\end{equation}
Among its physical roots, $r_{\rm I}$ is the inner boundary of the stable circular branch connected to large radius. Extra marginal roots, a negative $\mathcal{F}'$, or a nonpositive $D$ cannot be ignored. If that branch is absent, the present moving-disk prescription is unavailable, for the adopted emission law.

\begin{table}[!htbp]
\centering
\caption{Critical geometric data for the magnetic seeds of the nine parameter sets used in Figs.~\ref{fig:density_selected_nearcritical} and \ref{fig:energy_subcritical_2x3}. The mass $M$ is fixed and the nonlinear scale follows Eq.~\eqref{eq:mass_normalized_magnetic_constraint}. The columns give the geometric charge, outer seed horizon, selected unstable photon orbit, physical outer ISCO, and critical impact for the Maxwell test rays. All lengths are divided by $M$; $b_c$ defines the asymptotic critical screen radius. At the finite observer, its screen radius is given by Eq.~\eqref{eq:optical_finite_observer_screen}. Both copies have the same entries and identity impact matching. Values are rounded to three decimal places; no common throat radius is assigned.}
\label{tab:magnetic_optical_critical_data}
\begin{ruledtabular}
\begin{tabular}{ccccccc}
$\nu$ & $Q/Q_c$ & $Q/M$ & $r_h/M$ & $r_{\rm ph}/M$ & $r_{\rm I}/M$ & $b_c/M$\\
\hline
0.1 & 0.90 & 0.919 & 1.400 & 2.251 & 4.431 & 4.270\\
0.1 & 0.95 & 0.970 & 1.263 & 2.112 & 4.178 & 4.114\\
0.1 & 0.99 & 1.011 & 1.069 & 1.966 & 3.936 & 3.962\\
\hline
0.3 & 0.90 & 0.930 & 1.377 & 2.225 & 4.382 & 4.239\\
0.3 & 0.95 & 0.981 & 1.232 & 2.077 & 4.115 & 4.076\\
0.3 & 0.99 & 1.022 & 1.024 & 1.920 & 3.858 & 3.913\\
\hline
0.5 & 0.90 & 1.030 & 1.107 & 1.915 & 3.817 & 3.895\\
0.5 & 0.95 & 1.087 & 0.831 & 1.614 & 3.341 & 3.597\\
0.5 & 0.99 & 1.133 & 0.391 & 1.082 & 2.695 & 3.170\\
\end{tabular}
\end{ruledtabular}
\end{table}

Table~\ref{tab:magnetic_optical_critical_data} separates the geometric scales entering the shell and disk construction. The outer seed horizon bounds the static region that can be retained, while $r_{\rm ph}$ locates the maximum of the null potential accessible from infinity. The ISCO marks the inner edge of the stable circular timelike branch used for the emitter motion; it is not a second null critical orbit. In every listed case, $r_h<r_{\rm ph}<r_{\rm I}$, and the magnetic-field bound lies strictly inside $r_h$. Consequently the horizon, rather than that bound, sets the lower radial boundary for these exterior configurations. The horizon is removed by the throat excision and is not a horizon of the constructed wormhole. The roots were selected in the positive-lapse exterior, with negative null-potential curvature at $r_{\rm ph}$ and a stable circular branch extending outward from $r_{\rm I}$.

At each fixed $\nu$, increasing $Q/Q_c$ moves all three characteristic radii inward and decreases $b_c$ for the samples considered. For $\nu=0.5$, increasing the charge ratio from $0.90$ to $0.99$ changes $r_h/M$ from $1.107$ to $0.391$, $r_{\rm ph}/M$ from $1.915$ to $1.082$, and $r_{\rm I}/M$ from $3.817$ to $2.695$; the critical screen radius decreases from $3.895$ to $3.170$. The horizon therefore contracts more strongly than the exterior orbital scales over this interval. These variations follow the fixed-mass constraint: changing the charge also changes the nonlinear scale. They do not describe a charge scan at independently fixed $\beta_g$.

The inward shifts of the horizon and null orbit, and the decrease of the critical impact, have an analytic explanation on a continuous nondegenerate branch. At fixed $M$ and $\nu$, the constrained magnetic mass profile depends on the similarity variable $rM/Q^2$, up to a constant depending on $\nu$. Its positive magnetic energy density implies $(\partial_Q\mathcal F)_{r,M,\nu}=2(1-\mathcal F-r\mathcal F')/Q>0$. Thus the simple outer horizon moves inward because $\mathcal F'_h>0$. The corresponding implicit derivative of Eq.~\eqref{eq:photon_sphere_equation} is negative for an unstable orbit and an outward-decreasing magnetic density. At the null-potential maximum, the radius derivative of Eq.~\eqref{eq:optical_critical_impact} cancels, giving the particularly simple response
\begin{equation}
\left.\frac{\dd\ln(b_c/M)}{\dd Q}\right|_{M,\nu}
=\frac{3\mathcal F(r_{\rm ph})-1}{Q\mathcal F(r_{\rm ph})}<0.
\label{eq:optical_constrained_charge_response}
\end{equation}
The sign follows from the same positive-density identity evaluated at $r_{\rm ph}\mathcal F'=2\mathcal F$, which yields $\mathcal F(r_{\rm ph})<1/3$. This establishes the decrease of the geometric critical curve along the constrained charge family while the selected orbit remains admissible. The ISCO trend in the table is a numerical result for the stated samples; it is not inferred from the null-orbit response. At a degenerate horizon, the simple-root argument ceases to apply.

The $\nu=0.1$ and $0.3$ families remain close. At equal $Q/Q_c$, their photon radii, ISCO radii, and critical impacts differ by at most approximately $2.34\%$, $1.98\%$, and $1.22\%$, respectively, measured relative to $\nu=0.1$. Their common mass and closely spaced geometric charges give similar leading exterior terms in Eq.~\eqref{eq:rn_asymptotic}; nonlinear corrections distinguish the strong-field radii. The larger separation at $\nu=0.5$ also involves the change in $Q_c$ and in the constrained nonlinear scale. Comparing exponents at fixed normalized charge therefore compares different magnetic families, rather than varying $\nu$ while holding the dimensional charge and nonlinear scale fixed.

The asymptotic Reissner--Nordstr\"om form does not determine these strong-field scales by itself. For example, the last row has $Q/M\simeq1.133$: extrapolating the truncated Reissner--Nordstr\"om lapse to all radii would give neither a horizon nor a circular null orbit, since $Q>M$ and $9M^2-8Q^2<0$. The exact magnetic lapse nevertheless has the horizon and unstable orbit reported in the table. This contrast identifies the importance of retaining Eq.~\eqref{eq:seed_lapse} when solving the orbital conditions. Conversely, the formal $Q\to0$ limit at fixed $M$ and $\nu$, with $\beta_g\propto Q^6\to0$, approaches Schwarzschild at every fixed exterior radius. The limiting values are $(r_h,r_{\rm ph},r_{\rm I},b_c)/M=(2,3,6,3\sqrt{3})$. This is a limiting geometry; the endpoint is not an additional member of the adopted $Q>0$, $\beta_g>0$ branch.

The table does not choose $a_0$. Each standard-sector image requires a throat in its own interval $r_h<a_0<r_{\rm ph}$, together with Eq.~\eqref{eq:optical_barrier_sector}. In fact, a single throat cannot satisfy this interval for all nine entries: the largest outer horizon is $r_h/M\simeq1.400$, whereas the smallest photon radius is $r_{\rm ph}/M\simeq1.082$. The entries establish the available geometric sectors separately. Radial stability remains a further EOS-dependent condition, tested by Eq.~\eqref{eq:linearized_stability}; the existence of a horizon, photon barrier, and ISCO does not by itself establish a persistent shell.

Finally, Eq.~\eqref{eq:photon_throat_matching} makes the impact map the identity for every listed symmetric pair. Since the asymptotic lapse is unity, the observer-side and mapped second-side critical screen radii both equal $b_c$, independently of the allowed choice of $a_0$. There is consequently no finite interval of rays that cross the throat, reflect on the second side, and return because of unequal barriers. Bilateral emission still contributes through rays crossing once, while image brightness, ring peaks, and any central depression also depend on source support and frequency shifts. The critical data describe the minimally coupled test-ray geometry and do not substitute for nonlinear-electromagnetic characteristic cones or an emission calculation.

The specific timelike constants at $r_{\rm I}$ are
\begin{equation}
\mathcal{E}_{\rm I}=\frac{\mathcal{F}_{\rm I}}{\sqrt{D_{\rm I}}},\qquad
\ell_{{\rm I},i}=s_i\frac{r_{\rm I}^2\Omega_{\rm I}}{\sqrt{D_{\rm I}}}.
\label{eq:isco_emitter_constants}
\end{equation}
Here $\mathcal{E}_{\rm I}$ is dimensionless and $\ell_{{\rm I},i}$ is a signed length; both differ from the photon constants $E,L$ and the magnetic charge $q$. The inward continuation is
\begin{equation}
u_{{\rm pl},i}^{\mu}=\left(\frac{\mathcal{E}_{\rm I}}{\mathcal{F}},
-\sqrt{\mathcal{E}_{\rm I}^{2}-\mathcal{F}\left(1+\frac{\ell_{{\rm I},i}^{2}}{r^2}\right)},
0,\frac{\ell_{{\rm I},i}}{r^2}\right).
\label{eq:plunging_emitter_velocity}
\end{equation}
Its radial square root must remain real on the emitting interval, and the velocity is used only where $\mathcal{F}>0$. The branch conserves $-u_t=\mathcal{E}_{\rm I}$ and $u_{\varphi}=\ell_{{\rm I},i}$ and approaches the circular ISCO velocity continuously. An exactly marginal circular particle does not spontaneously plunge; this branch describes its inward separatrix after an inward perturbation. The flows are prescribed independently in the two exteriors, inward toward the shell in their own areal charts. Photon transparency does not impose a continuation of a gas worldline or a hydrodynamic mass flux across the shell. Any such coupling would require additional matter matching. No shell luminosity is included.

\subsection{Emission model C and its physical support}\label{subsec:optical_emission}

We adopt model C as the common radial emission law in both exteriors. This arctangent law follows the static disk prescription of Ref.~\cite{ChenYang2025}, with its radial argument expressed here in units of $M$. The prescribed bolometric crossing intensity is measured in each emitter rest frame, with reference intensity $I_0>0$. The mass scale $M$ established in \cref{sec:seed} supplies the common radial width, and the seed horizon fixes the normalization of the dimensionless profile:
\begin{equation}
\mathcal J_{\rm C}(r)=
\begin{cases}
\dfrac{\dfrac{\pi}{2}-\arctan\!\left(\dfrac{r-r_{\rm I}+M}{M}\right)}
{\dfrac{\pi}{2}-\arctan\!\left(\dfrac{r_h-r_{\rm I}+M}{M}\right)},&r>r_h,\\
0,&r\leq r_h.
\end{cases}
\label{eq:optical_emission_c}
\end{equation}
This arctangent profile approaches unity at the formal horizon edge, is positive and decreasing outside it, and has an $r^{-1}$ tail. It can illuminate the plunging region below $r_{\rm I}$, including radii below $r_{\rm ph}$. The normalization retains the seed value $r_h$ even though the wormhole emitting support starts at the throat; replacing it by $a_0$ would change the prescribed profile. The adopted moving source also requires an admitted physical ISCO and the specified velocity domain.

The local source intensity on either side is
\begin{equation}
I_{{\rm em},i}(r)=I_0\xi_i\mathcal J_{\rm C}(r)\mathbf 1_{\mathcal S_i}(r).
\label{eq:optical_side_sources}
\end{equation}
Here $i=1,2$ labels the emitting exterior, with $\xi_1=1$ and $\xi_2>0$, and $\mathbf 1_{\mathcal S_i}$ enforces its physical support, including $r\geq a_0$, the common finite outer radius $r_{{\rm out},i}=10^3M$, and a real future-directed timelike emitter velocity. For moving emitters below the ISCO, the emitting region is restricted to the inward branch continuously connected to it. A negative velocity radicand marks a forbidden radial interval. If that branch turns before reaching the shell, the turning point bounds the moving-source support. The static control instead requires only a positive lapse and its prescribed radial support. For a matched black hole, the corresponding geometric support is $r>r_h$, with its own admitted velocity interval, rather than $r\geq a_0$. Removing an emitting interval from the wormhole does not redefine the formal source law.

The same profile and radial width apply to both universes. Relative intensity $\xi_2$, outer radii and rotation senses remain explicit source parameters; equal illumination requires $\xi_2=1$ together with equal outer supports and aligned gas velocities. Thus a common emission law does not by itself fix the relative brightness or motion of the two disks.

The prescribed intensity is phenomenological and locally isotropic; it is not a temperature-derived emissivity or a hydrodynamic disk solution. No additional path-length or limb-darkening factor is imposed. A finite-thickness volume source would require a line-of-sight transfer integral, particularly near edge-on viewing. Comparisons use a common $I_0$ and a declared display stretch; independent peak normalization of each image would conceal relative source brightness.

\begin{figure}[!htbp]
\centering
\includegraphics[width=\textwidth]{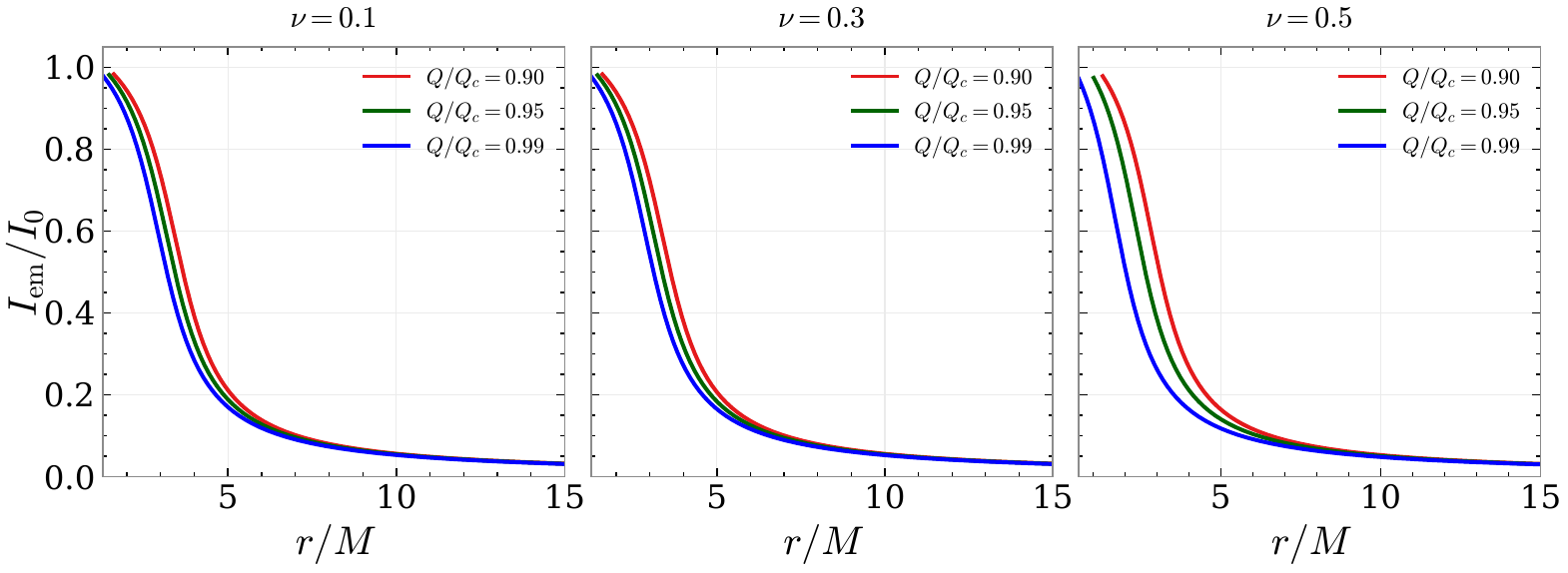}
\caption{Normalized rest-frame bolometric emission $I_{\rm em}/I_0$ of model C versus $r/M$. Columns correspond to $\nu=0.1$, $0.3$, and $0.5$; red, green, and blue curves represent $Q/Q_c=0.90$, $0.95$, and $0.99$, respectively. The fixed-mass magnetic parameters are those of Table~\ref{tab:magnetic_optical_critical_data}. Both exteriors have unit relative intensity, the same seed-horizon normalization and width $M$, and source support from $a_0=r_h+0.25(r_{\rm ph}-r_h)$ to $10^3M$. Each column begins at its smallest $a_0$ and displays radii through $15M$. No gravitational or Doppler transfer factor is included.}
\label{fig:emission_C_subcritical_1x3}
\end{figure}

Figure~\ref{fig:emission_C_subcritical_1x3} isolates the radial source prescription from photon propagation. On the common displayed support of each column, increasing $Q/Q_c$ shifts the decline inward and lowers the local intensity at a fixed radius. This dependence enters Eq.~\eqref{eq:optical_emission_c} through the geometric radii $r_h$ and $r_{\rm I}$, with the magnetic scale constrained by Eq.~\eqref{eq:mass_normalized_magnetic_constraint}; it does not represent a gravitational or Doppler attenuation of the emitted radiation. Differentiation of the prescribed law gives a strictly negative slope whose magnitude is largest at $r=r_{\rm I}-M$, provided that radius lies in the source support. All nine configurations satisfy this condition. The inward displacement of the ISCO therefore moves the steepest part of the profile inward while its width remains fixed by $M$. The larger ISCO displacement for $\nu=0.5$ explains the stronger charge separation in the third column.

The close appearance of the first two columns follows from their nearby horizon and ISCO radii in Table~\ref{tab:magnetic_optical_critical_data}, rather than from an additional adjustment of the emission law. At equal $Q/Q_c$, their normalized profiles differ by less than $0.028$ throughout their common displayed support. The different curve beginnings have a separate geometric origin: the prescribed $a_0$ follows both the shrinking horizon and the changing horizon--photon-sphere gap. In the third column, $a_0/M$ decreases from approximately $1.309$ to $0.564$ as $Q/Q_c$ increases from $0.90$ to $0.99$. These are distinct retained source boundaries, not zeros of the emission or numerical clipping. The initial intensities remain finite, between approximately $0.972I_0$ and $0.983I_0$, because the normalization refers to the excised seed horizon rather than to the throat. This choice of $a_0$ specifies a geometric optical comparison and does not establish radial stability under any of the surface equations of state.

At larger radii the profiles flatten and their absolute differences decrease. Expanding Eq.~\eqref{eq:optical_emission_c} gives an inverse-radius tail whose coefficient is $M$ divided by the horizon normalization denominator in that equation. This coefficient retains its dependence on $r_h$ and $r_{\rm I}$, so convergence toward zero does not imply identical tails or a ratio approaching unity. For $\nu=0.5$, the charge ordering reverses weakly beyond the displayed interval, so the ordering in the figure is not a global ordering of the source law. At the adopted outer edge $r=10^3M$, the nine configurations have local intensities between $3.74\times10^{-4}I_0$ and $4.03\times10^{-4}I_0$. The formal infinite-radius expansion explains the flattening, whereas the physical source has finite support. Its sharp outer truncation remains an imposed boundary condition: when a disk crossing leaves that support, its nonzero contribution to Eq.~\eqref{eq:optical_bolometric_sum} disappears. The larger outer radius suppresses this edge effect through the inverse-radius emission tail without making the finite-support transfer mathematically continuous. It does not alter the curves displayed through $15M$ or the geometric critical parameters in Table~\ref{tab:magnetic_optical_critical_data}.

The figure consequently fixes the intrinsic brightness supplied at an admitted disk crossing. It does not determine a bright-ring position, a screen intensity maximum, or the total contrast between the wormhole and its black-hole control. Those observables also depend on the crossing radii, emitter velocities, frequency shifts and side-resolved ray contributions in Eq.~\eqref{eq:optical_bolometric_sum}. Equal local profiles on the two sides therefore specify equal source prescriptions without ensuring equal observed contributions.

\subsection{Frequency shifts and bolometric transfer}\label{subsec:optical_transfer}

For a future-directed photon, the local frequency ratio follows directly from the observer and emitter contractions:
\begin{equation}
g=\frac{-k_{\mu}u_{\rm obs}^{\mu}}{-k_{\mu}u_{\rm em}^{\mu}}.
\label{eq:optical_frequency_ratio}
\end{equation}
For the static observer, $u_{\rm obs}^{t}=\mathcal F_o^{-1/2}$ and the numerator is $E/\sqrt{\mathcal F_o}$. It tends to $E$ at infinity. With $k_t=-E$ and the normalized covariant spatial momenta $\widehat k_r=k_r/E$, $\widehat k_{\varphi}=k_{\varphi}/E$, the equality $E_1=E_2$ makes the frequency ratio for an emitter on either side
\begin{equation}
g_i=\left\{\sqrt{\mathcal F_o}\left[u_i^t-u_i^r\widehat k_{r,i}-u_i^{\varphi}\widehat k_{\varphi,i}\right]\right\}^{-1}.
\label{eq:optical_moving_redshift}
\end{equation}
The denominator must be positive on an accepted emission event. All momenta in this expression are future-directed, even when the numerical trajectory is traced backward from the observer. This convention is essential for the radial Doppler term of an inward emitter.

Conservation of specific intensity divided by the cube of frequency along a transparent vacuum ray gives $I_{{\rm obs},\omega_{\rm ph}}=g^3 I_{{\rm em},\omega_{\rm ph}/g}$ for spectral angular frequency $\omega_{\rm ph}$. Integrating over observed frequency supplies one further factor of $g$. The bolometric screen intensity is therefore
\begin{equation}
I_{\rm obs}(b,\chi;\iota)
=\sum_{i=1}^2\sum_{m\in\mathcal H_i(b,\chi;\iota)}
g_{i,m}^4 I_{{\rm em},i}(r_{i,m})
=I_{\rm WH,1}+I_{\rm WH,2}.
\label{eq:optical_bolometric_sum}
\end{equation}
Let $\iota$ denote the angle between the viewing direction and the aligned disk normal, with $0\leq\iota<\pi/2$, and let $\chi$ be screen azimuth measured from the $X$ axis. The exactly edge-on limit is excluded for the adopted zero-thickness disks. At finite $r_o$, let $\alpha_o$ be the local angular separation from the inward viewing direction, $K_o=r_o/\sqrt{\mathcal F_o}$, and $\rho_{\rm scr}$ the radius on the stereographic screen. The local viewing angle is related to the conserved impact parameter by
\begin{equation}
\sin\alpha_o=\frac{b}{K_o}.
\label{eq:optical_local_viewing_angle}
\end{equation}
The stereographic projection converts this angle into the screen radius:
\begin{equation}
\rho_{\rm scr}=2K_o\tan\frac{\alpha_o}{2}
=\frac{2b}{1+\sqrt{1-b^2/K_o^2}}.
\label{eq:optical_finite_observer_screen}
\end{equation}
Conversely, a screen radius determines the impact parameter through
\begin{equation}
b(\rho_{\rm scr})=\frac{\rho_{\rm scr}}{1+\rho_{\rm scr}^2/(4K_o^2)}.
\label{eq:optical_inverse_observer_screen}
\end{equation}
Thus $X=\rho_{\rm scr}\cos\chi$ and $Y=\rho_{\rm scr}\sin\chi$, tending to the impact-plane coordinates at infinity. Across the displayed square $|X|,|Y|\leq7M$, the relative screen-radius correction is below $3\times10^{-9}$. The physical intensity normalization $I_0$ and apertures are held fixed between the static and moving comparisons. Each face-on figure uses a common color reference across its four rows; the inclined maps share a reference within each inclination row. The display stretches and references are declared in the captions, so equal color levels across different references need not represent the same $I_{\rm obs}/I_0$.

The set $\mathcal H_i$ contains disk intersections in exterior $i$, with its own source and velocity support imposed at each event. The side contributions include $I_0\xi_i$ from \cref{eq:optical_side_sources}. Their indices follow the global geometric order; they are not restarted at the shell. An event on the joining surface is counted only once, with no independent shell-emission term. Contralateral emission contributes even though there is no reflected returned family. Multiple intersections add because the disk is transparent. There is no separate lensing magnification factor multiplying the specific intensity at a fixed screen point; lensing enters through the transfer map and through the apparent solid angle in a flux integral.

The first, second, and subsequent global intersections resolve the direct and increasingly lensed contributions. A finite intersection cutoff approximates the contribution from both sides. Convergence is assessed by increasing the retained order and the radial and angular resolution for each side contribution and aperture observable. Such a truncation does not establish convergence arbitrarily near $b_c$, where the angular advance diverges. Quantitative peak and ring-contrast estimates remain conditional on bounds for the higher-order tails and finite-resolution effects. The formal sum in \cref{eq:optical_bolometric_sum} defines the observable independently of that provisional truncation.

\subsection{Imaging results}\label{subsec:optical_imaging_results}

We compare the bilateral wormhole with its matched seed black hole, using model C throughout. Face-on viewing includes a static-emitter control and the circular/plunging source; inclined viewing retains the latter. A common observer, emission law and set of image observables make the role of the retained source domain explicit: the wormhole receives light from both exteriors, whereas the black-hole control can also emit from radii removed by the throat excision.

\subsubsection{Comparison setup and image observables}\label{subsec:optical_observables}

The seed black hole provides a matched control with the same lapse, magnetic parameters, constrained mass, observer, inclination and model-C source prescription. Its rays are absorbed at the horizon. The BH disk has the same $I_0$, emitter prescription and outer radius as the disk in $\mathcal M_1$, including the rotation sense when motion is present; it has no counterpart of the second universe. Comparing different seed parameters can be informative, but changes both geometry and source radii and does not isolate the effect of the throat.

For that control, $\mathcal H_{\rm BH}$ denotes the disk-crossing events on the absorbing BH ray, $g_{{\rm BH},m}$ is evaluated using \cref{eq:optical_frequency_ratio} with the same gas prescription, and $I_{{\rm em},\rm BH}$ uses \cref{eq:optical_side_sources} with $\xi_1=1$ and the BH radial support. Inclination enters through the crossing locations and the observer--emitter frequency contractions.

The basic observables are $I_{\rm obs}/I_0$ on the $(X/M,Y/M)$ screen, individual intersection-order contributions, and the locations and widths of resolved intensity features. Two useful one-dimensional diagnostics are
\begin{equation}
\mathcal{P}_{r}(b;\iota)=\int_0^{2\pi}I_{\rm obs}(b,\chi;\iota)\dd\chi,
\label{eq:optical_radial_integrated_profile}
\end{equation}
and
\begin{equation}
\mathcal{P}_{\chi}(\chi;\iota)=\int_0^{b_{\max}}I_{\rm obs}(b,\chi;\iota)\dd b.
\label{eq:optical_azimuth_integrated_profile}
\end{equation}
The first has intensity dimensions and resolves radial structure; the second has intensity times length dimensions and resolves azimuthal asymmetry for a specified screen cutoff. Neither is a total flux, because the screen-area Jacobian is absent. For a distant observer at areal distance $D_o$, the flux within that screen window is
\begin{equation}
\mathscr{F}_{\rm obs}\simeq\frac{1}{D_o^2}\int_0^{b_{\max}}\!\int_0^{2\pi}
I_{\rm obs}(b,\chi;\iota)b\,\dd\chi\dd b.
\label{eq:optical_area_weighted_flux}
\end{equation}
The approximation is the distant, small-angle relation between impact-plane area and observed solid angle, with $D_o\simeq r_o$. The finite-observer corrections are of relative order $M/r_o$ from the lapse and $b^2/r_o^2$ from the angular projection; no aperture flux is inferred from a color-stretched map. The aperture flux depends on the chosen image window and the finite outer radius of each disk. The $r^{-1}$ tail of model C requires an explicit outer cutoff for a finite source-luminosity comparison. Each side is therefore integrated over its declared source domain and the same image aperture is used in every control. Side-resolved profiles and aperture fluxes obey the same additive decomposition as \cref{eq:optical_bolometric_sum}.

In the standard sector \cref{eq:optical_barrier_sector}, supercritical rays turn outside the shell, so their source paths and weights agree in the WH and matched BH. Subcritical backward rays follow the same $\mathcal M_1$ segment down to $a_0$ and then either continue into $\mathcal M_2$ or proceed toward the BH horizon. They do not reflect back in the symmetric geometry. Let $I_{\rm BH,ret}$ be the BH contribution from the retained observer-side interval $r\geq a_0$, and define the excised-source contribution by
\begin{equation}
I_{\rm BH,exc}=
\sum_{m\in\mathcal H_{\rm BH}:\,r_h<r_m<a_0}
g_{{\rm BH},m}^{4} I_{{\rm em},\rm BH}(r_m)\geq0.
\label{eq:optical_bh_excised_source}
\end{equation}
The sum includes only events within the BH source and timelike velocity support. With equal observer-side support on the common interval, the decomposition is
\begin{equation}
\begin{aligned}
I_{\rm WH,1}&=I_{\rm BH,ret},\\
I_{\rm BH}&=I_{\rm BH,ret}+I_{\rm BH,exc},\\
I_{\rm WH}-I_{\rm BH}
&=I_{\rm WH,2}-I_{\rm BH,exc}.
\end{aligned}
\label{eq:bilateral_wh_bh_difference}
\end{equation}
These identities hold pixel by pixel, for either inclination, away from the limiting critical ray and subject to the stated barrier and common-source conditions. They distinguish radiation transmitted from the opposite universe from BH radiation removed by the cut. Side-resolved intersections determine the emitting exterior responsible for a bright feature; morphology alone does not identify its origin.

The seed-BH disk may emit at $r_h<r<a_0$, so its excised contribution can compete with the added light from $\mathcal M_2$. The full WH--BH difference therefore has no prescribed sign. A separate support-matched BH control, whose emission is deliberately cut off at $a_0$ while its horizon and geometry remain unchanged, removes the source-excision term and isolates the transmitted intensity. This truncated-source control has different radial support from the untruncated seed disk. Supercritical pixels receive neither contralateral light nor the excised inner contribution in the standard sector; their equality does not imply equality of the complete images.

All comparisons use model C. Variations in $\xi_2$, outer support or rotation are separate controls at fixed emission law and geometry. A source-induced ring or brightness change is not by itself a unique geometric signature of a wormhole.

Both viewing configurations use the source domains and seed parameters established in \cref{sec:seed}. The same moving-source support is used face-on and inclined; the static face-on control does not impose a plunging-velocity restriction. The physical-orbit, timelike-emitter and shell-stability requirements remain separate conditions. Quantitative ring locations, amplitudes and contralateral fractions depend on the numerically evaluated transfer functions and on convergence in spatial resolution and retained intersection order.

\subsubsection{Face-on images and bilateral emission}\label{subsec:optical_face_on}

\begin{figure}[!htbp]
\centering
\includegraphics[width=\textwidth,height=0.78\textheight,keepaspectratio]{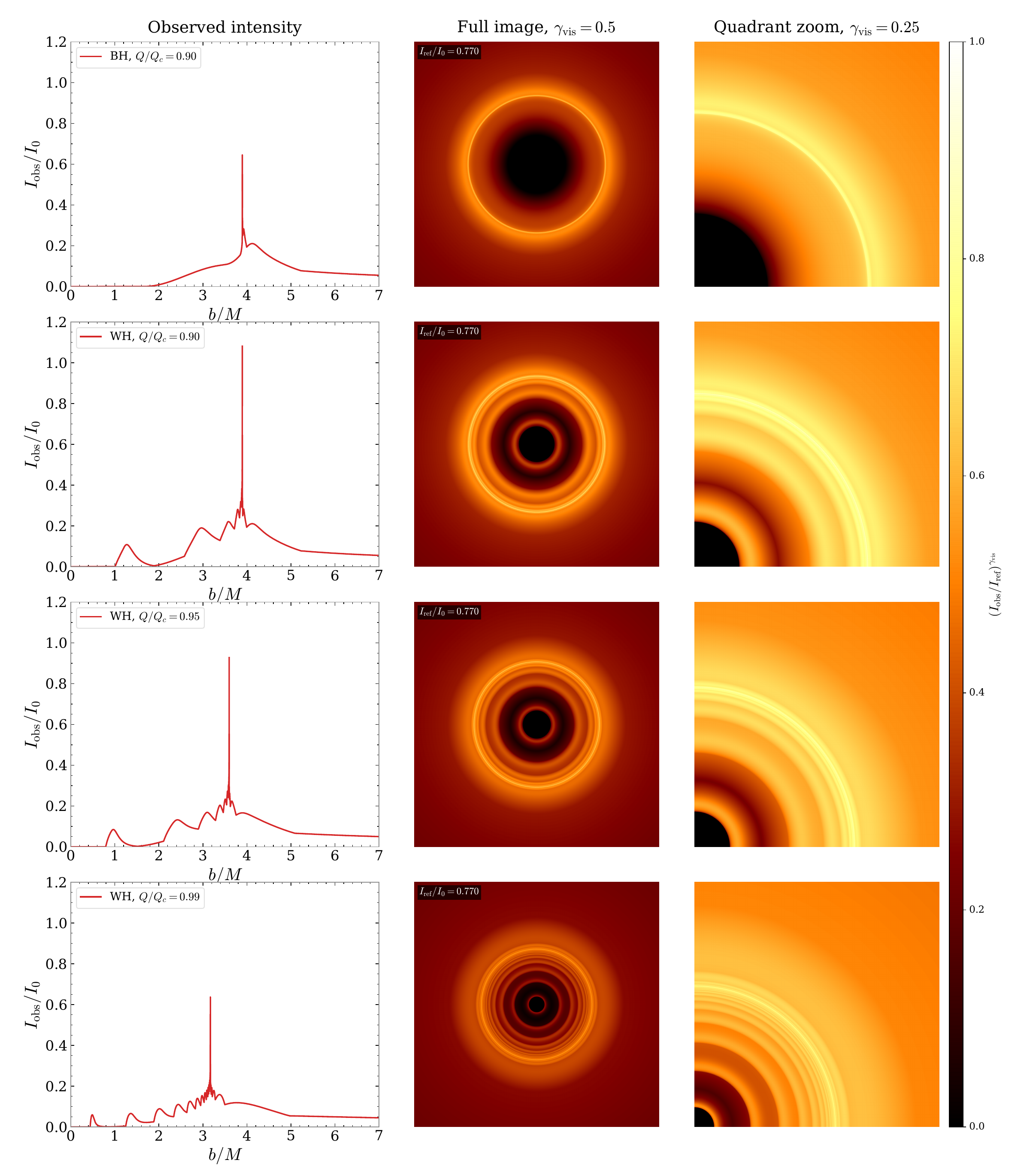}
\caption{Face-on bolometric intensity of static model-C emitters at $\nu=0.5$, observed by a static observer at $r_o=10^5M$. Rows show the magnetic seed BH at $Q/Q_c=0.90$ and bilateral WHs at $Q/Q_c=0.90$, $0.95$ and $0.99$. Columns show $I_{\rm obs}/I_0$ for $0\leq b/M\leq7$, the full screen $-7\leq X/M,Y/M\leq7$, and its first-quadrant enlargement $0\leq X/M,Y/M\leq5.5$. The curve window is $0\leq I_{\rm obs}/I_0\leq1.2$. Both WH disks have unit relative amplitude, the throat is $a_0=r_h+0.25(r_{\rm ph}-r_h)$ and the outer source boundary is $10^3M$. The maps display $(I_{\rm obs}/I_{\rm ref})^{\gamma_{\rm vis}}$, with $\gamma_{\rm vis}=1/2$ for the full screen and $1/4$ for the enlargement. This dimensionless contrast exponent is distinct from the electromagnetic coupling $\gamma$. One finite-pixel reference $I_{\rm ref}/I_0=0.7703448$ is used across all four rows.}
\label{fig:face_on_static_C}
\end{figure}

\begin{figure}[!htbp]
\centering
\includegraphics[width=\textwidth,height=0.78\textheight,keepaspectratio]{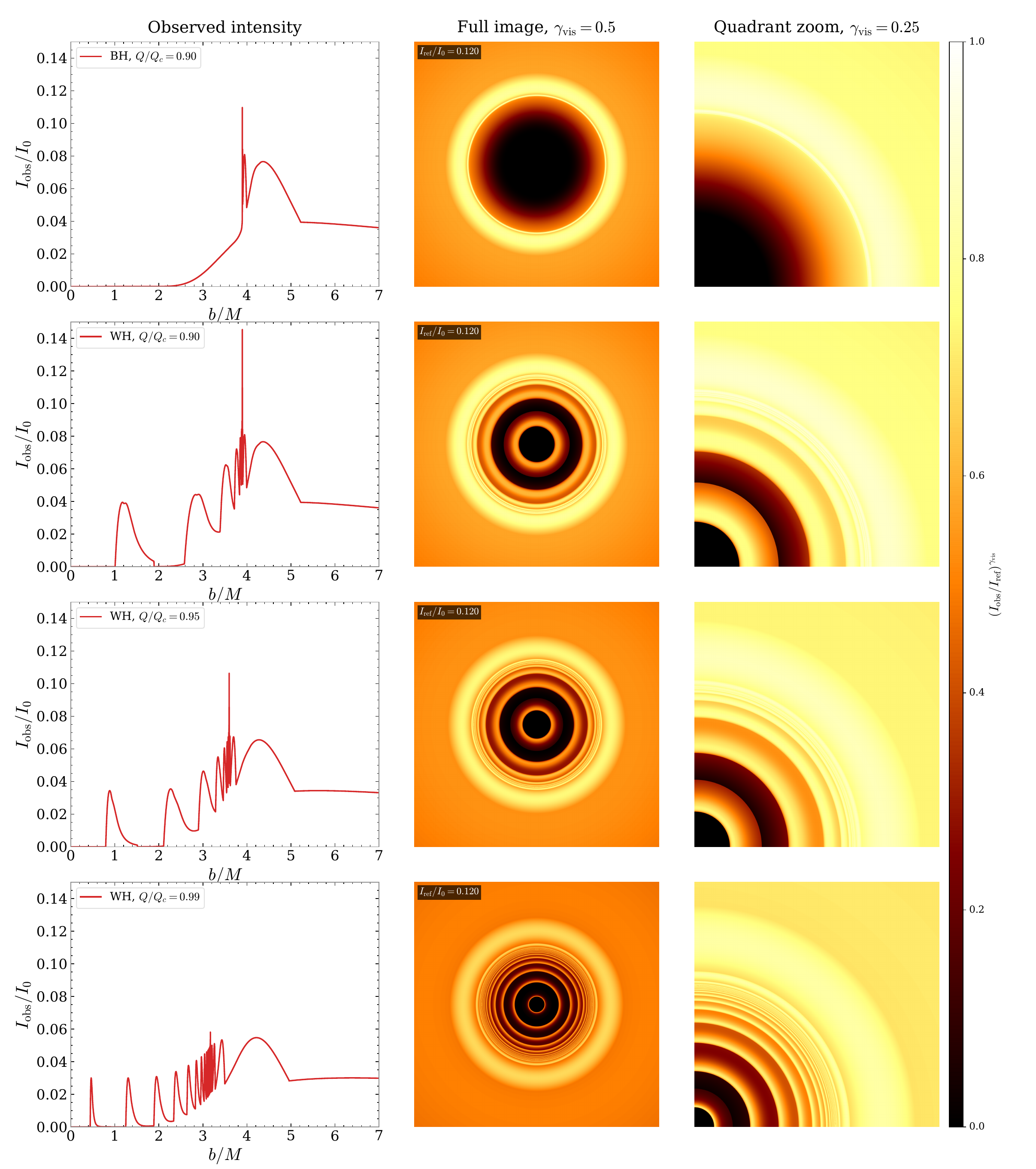}
\caption{Face-on model-C intensity for moving emitters at $\nu=0.5$. The observer, magnetic seeds, row order, source amplitudes, throat prescription, outer boundary and screen windows are those of Fig.~\ref{fig:face_on_static_C}. Emitters follow circular timelike geodesics outside the ISCO and the inward ISCO separatrix inside it. The left-column window is $0\leq I_{\rm obs}/I_0\leq0.15$. The full and first-quadrant maps use $\gamma_{\rm vis}=1/2$ and $1/4$, with a common finite-pixel reference $I_{\rm ref}/I_0=0.1196853$ across the four rows. This color reference differs from that of the static figure; the radial curves retain the same physical normalization $I_0$.}
\label{fig:face_on_moving_C}
\end{figure}

The face-on comparison uses $\iota=0$ and first separates the effects of ray geometry and gravitational frequency shifts from those of emitter motion. Figures~\ref{fig:face_on_static_C} and \ref{fig:face_on_moving_C} use the same magnetic seeds, throat prescription, model-C emission and static observer. In each figure, the first row is the seed BH at $Q/Q_c=0.90$, followed by bilateral WHs at $Q/Q_c=0.90$, $0.95$ and $0.99$, all at $\nu=0.5$. Only the equal-charge BH--WH pair is a matched geometric control. The other WH rows vary the constrained magnetic family, including the geometric radii entering the source law. The throat is set to $a_0=r_h+0.25(r_{\rm ph}-r_h)$ separately for each seed; these optical choices satisfy the retained geometric sector, without selecting an EOS-stable shell.

For viewing along the disk normal, the photon orbital plane has zero angular momentum about the disk axis, $\widehat k_{\varphi}=0$. Measured from the observer, the accumulated backward angles at disk crossings are
\begin{equation}
\psi_m=\frac{\pi}{2}+(m-1)\pi,\qquad m=1,2,\ldots.
\label{eq:face_on_disk_targets}
\end{equation}
A target contributes only if the actual path reaches it in either exterior and its radius lies within the source support. Inverting Eq.~\eqref{eq:optical_angle_integral} on the incoming, outgoing or transmitted branch determines the crossing radii and future radial signs. Those geometric events are common to the two emitter prescriptions; their weights are evaluated with the corresponding velocity.

The aligned disks give an axisymmetric screen intensity,
\begin{equation}
I_{\rm obs}(X,Y;0)=I_{\rm obs}\!\left(b(\rho_{\rm scr});0\right),\qquad
\rho_{\rm scr}=\sqrt{X^2+Y^2},
\label{eq:face_on_screen_map}
\end{equation}
where $b(\rho_{\rm scr})$ is defined by Eq.~\eqref{eq:optical_inverse_observer_screen}. Each side contribution is also axisymmetric, even for unequal source amplitudes. The critical circle has radius $\rho_{\rm scr}(b_c)$; the first illuminated radius and the intensity maximum need not lie on it. The first-quadrant enlargement samples the same intensity as the full screen, while changing its window and display stretch. The radial diagnostic in Eq.~\eqref{eq:optical_radial_integrated_profile} equals $2\pi I_{\rm obs}(b;0)$.

\paragraph{Static emitters.}
Substituting Eq.~\eqref{eq:static_emitter_velocity} into the frequency contraction gives
\begin{equation}
g_{\rm stat}(r)=\sqrt{\frac{\mathcal F(r)}{\mathcal F_o}},\qquad
g_{\rm stat}^{4}(r)=\left[\frac{\mathcal F(r)}{\mathcal F_o}\right]^2.
\label{eq:static_face_on_weight}
\end{equation}
There is no velocity-dependent Doppler term, and a given radius has the same weight on every radial leg. Gravitational attenuation still competes with the inward increase of the intrinsic emission, while multiple intersections add through Eq.~\eqref{eq:optical_bolometric_sum}.

Figure~\ref{fig:face_on_static_C} shows narrow intensity enhancements around the geometric critical circle, embedded in a broader direct-emission component. Near $b_c$, rays spend a longer angular interval close to the unstable photon orbit and intersect the transparent disk repeatedly. The additional positive contributions can dominate the intensity even though each crossing is gravitationally attenuated. This mechanism associates the narrow annular structure with strongly deflected rays, without identifying every visible maximum with the exact critical orbit. That limiting orbit has infinite angular advance and is excluded from the finite ray sample. Peak heights and the apparent widths of the narrowest features therefore depend on the resolved distance from $b_c$, the intersection truncation and the pixel sampling; they are not continuum peak estimates. In particular, the reference used for the maps is their finite-pixel maximum, not the maximum of the more finely sampled radial curves.

The first two rows isolate the effect of replacing the BH interior with a transparent throat. For $b>b_c$, their curves agree because the ray turns before reaching the excision, preserving all source crossings and weights. Below $b_c$, Eq.~\eqref{eq:bilateral_wh_bh_difference} separates the transmitted emission from the BH light lost through source excision. In these samples, the WH illumination begins at approximately $b/M=1.01$, whereas the BH begins at approximately $1.64$. A smaller dark central region is thus compatible with the absence of a WH horizon, but it is not a measurement of the throat radius: the ray must accumulate enough angle to reach an admitted equatorial crossing. At $b=0$ no such crossing occurs for the zero-thickness source. A dark center in this model does not by itself distinguish horizon absorption from the geometry of a transparent disk image. The transmitted contribution brightens part of the subcritical domain in the displayed matched pair; its sign relative to an untruncated BH is still governed by the competition in Eq.~\eqref{eq:bilateral_wh_bh_difference}.

The successive WH rows move the critical annuli inward, following the decrease of $b_c/M$ from $3.895$ to $3.597$ and $3.170$ in Table~\ref{tab:magnetic_optical_critical_data}. Their illuminated inner boundaries also move inward, to approximately $b/M=0.79$ and $0.44$. These two shifts have different definitions: the first follows the null-potential maximum, while the second follows the first admitted disk crossing. Both arise from the changing exact geometry and retained source domain. The accompanying brightness changes also include the displacement of the ISCO-centered source profile and its seed-horizon normalization. They cannot be attributed to charge alone at independently fixed nonlinear scale or to an added throat-emission term.

The diffuse exterior emission decreases much more slowly than the narrow critical features. Far from the central region, direct crossings dominate and the static transfer tends to $\mathcal F_o^{-2}$, with the source retaining its inverse-radius tail before the finite boundary. The enlargements reveal faint structure through their stronger display stretch, rather than through a different physical intensity. Their common reference makes relative row brightness meaningful within this figure, but a color-stretched area is not an integrated-flux observable.

\paragraph{Moving emitters.}
For circular gas, the face-on frequency ratio instead becomes
\begin{equation}
g_{\rm circ}(r;0)=\sqrt{\frac{D(r)}{\mathcal F_o}},\qquad
g_{\rm circ}^{4}(r;0)=\frac{[\mathcal F(r)-r\mathcal F'(r)/2]^2}{\mathcal F_o^2}.
\label{eq:moving_face_on_circular_weight}
\end{equation}
Viewing along the disk normal removes the approaching--receding rotational asymmetry but retains the transverse time dilation. At any shared circular crossing with $\mathcal F'>0$ and $D>0$, the ratio of moving to static bolometric weights is $g_{\rm circ}^4/g_{\rm stat}^4=(D/\mathcal F)^2<1$. This supplies an analytic suppression criterion independent of the local emission amplitude.

For the inward separatrix, let $s_r=\operatorname{sgn}(k^r)$ be the future radial photon sign in its emitting exterior. Equations~\eqref{eq:optical_null_potential} and \eqref{eq:optical_moving_redshift} give
\begin{equation}
g_{{\rm pl},i}(r,b;0)=\frac{\mathcal F(r)}{\sqrt{\mathcal F_o}\,[\mathcal E_{\rm I}
-s_r u^r_{{\rm pl},i}\sqrt{1-b^2\mathcal U(r)}]}.
\label{eq:moving_face_on_plunge_weight}
\end{equation}
The radial Doppler term survives face-on viewing. Since $u^r_{{\rm pl},i}<0$, an outward future photon has a larger denominator than an inward one at the same radius and impact. The distinct weights on the observer-side and transmitted legs must therefore be retained separately. Circular suppression does not establish a universal inequality for all plunging contributions.

Figure~\ref{fig:face_on_moving_C} preserves the critical radii, axisymmetry and illuminated inner boundaries of Fig.~\ref{fig:face_on_static_C}, while substantially reducing the intensities around the annular enhancements. The velocity changes the frequency weights, not the null trajectories or geometric crossing targets. The selected moving branches remain admitted over the common source domain for these parameter sets, so the unchanged inner boundaries are expected. The critical features consequently retain their geometric charge ordering, while their amplitudes and contrast change. Their greater prominence relative to a diffuse component in a stretched map should not be interpreted as a larger absolute intensity: the two figures have different curve windows and map references. Absolute comparisons must use $I_{\rm obs}/I_0$ in the left columns.

The analytic circular suppression and the leg-dependent plunging shift account for different parts of this reduction. For example, at $Q/Q_c=0.99$ and $b/M=2$, the total intensity falls from about $0.0884I_0$ for static emitters to $0.0225I_0$ for moving emitters. The observer-side and opposite-side contributions both decrease at that screen location, but by different factors because their crossing radii and future radial signs differ. Thus the reduction is not reproduced by multiplying the entire static image by one constant. Nor can a bright subcritical feature be assigned to one exterior solely from its appearance. The side decomposition and the BH-excision term remain necessary in both emitter models. At the matched charge $Q/Q_c=0.90$, supercritical BH and WH intensities still agree separately for each velocity prescription.

The nearly flat outer part of the last moving row has a distinct explanation. Between $b/M=6$ and $7$, its intensity changes only from approximately $0.02996I_0$ to $0.02986I_0$, whereas the static profile decreases from $0.04922I_0$ to $0.04459I_0$. These rays have a single observer-side crossing outside the ISCO, so the relevant velocity effect is circular time dilation, not the inward plunge or transmission through the throat. Over the corresponding moderate radii, the increasing factor $D^2/\mathcal F_o^2$ partly compensates the decreasing model-C emission. For this seed, the product $g_{\rm circ}^4 I_{\rm em}/I_0$ is approximately $0.02983$ at $r=5M$ and $0.02994$ at $r=6M$. This compensation explains the broad shoulder without requiring a constant intrinsic emissivity or a numerical step.

The shoulder is not a nonzero asymptotic intensity. From Eq.~\eqref{eq:rn_asymptotic}, $\mathcal F=1-2M/r+O(r^{-2})$ and $D=1-3M/r+O(r^{-2})$, so the circular/static weight ratio tends to unity as $(D/\mathcal F)^2=1-2M/r+O(r^{-2})$. Combined with the model-C inverse-radius tail and the asymptotically direct transfer, both observed profiles decay as $1/b$ in the distant intermediate region $M\ll b\ll r_{\rm out},r_o$. For $Q/Q_c=0.99$, the moving and static intensities are approximately $0.01190I_0$ and $0.01278I_0$ at $b=30M$, and $0.003887I_0$ and $0.003967I_0$ at $b=100M$. Their decreasing separation supports the weak-field limit, without extending the finite source beyond $10^3M$. At its edge, the disappearance of an admitted crossing is still a source-boundary effect; it is not the origin of the shoulder displayed at $b/M\leq7$.

The static face-on control therefore identifies the geometric crossing structure and gravitational attenuation, while the moving comparison exposes transverse and radial frequency shifts at fixed source law. For inclined viewing we retain only the moving prescription, which additionally couples the rotation to a nonzero line-of-sight photon momentum. This choice carries the same specified accretion-like test flow through the inclination study, allowing the orientation dependence to be examined without simultaneously changing the emitter model. An inclined static control is mathematically possible, but the present extension focuses on the azimuthal Doppler contrast and its interplay with bilateral transmission; the face-on static analysis already provides the reference for separating source motion from ray geometry.

\subsubsection{Inclined images and bilateral emission}\label{subsec:optical_inclined}

For $0<\iota<\pi/2$, we retain the circular/plunging emitters of Fig.~\ref{fig:face_on_moving_C}. The bilateral wormhole and its seed-BH control are imaged at the same inclination, aperture and physical intensity scale. To specify the orientation and determine disk intersections, write a reference orbital-plane ray as $r(\cos\psi,\sin\psi,0)$, with $\psi$ increasing on the backward trajectory from the observer. Its position in disk coordinates is obtained using
\begin{equation}
\mathsf{R}(\iota,\chi)=
\begin{pmatrix}
\sin\iota&-\cos\iota\sin\chi&-\cos\iota\cos\chi\\
0&\cos\chi&-\sin\chi\\
\cos\iota&\sin\iota\sin\chi&\sin\iota\cos\chi
\end{pmatrix}.
\label{eq:inclined_ray_rotation}
\end{equation}
This orthogonal rotation fixes both the disk-intersection convention and the sign of the azimuthal photon momentum. The height above the disk is $z/r=\cos\iota\cos\psi+\sin\iota\sin\chi\sin\psi$. Its first positive zero is
\begin{equation}
\psi_c(\iota,\chi)=\operatorname{atan2}\!\left(\cos\iota,-\sin\iota\sin\chi\right).
\label{eq:inclined_first_crossing_angle}
\end{equation}
Since $\cos\iota>0$, this choice lies between zero and $\pi$ and removes the branch ambiguity of an ordinary arctangent. Subsequent disk targets are $\psi_m=\psi_c+(m-1)\pi$. They are tested against the actual angular extent of every allowed ray segment, preserving global intersection order across the throat.

The normalized covariant momenta entering \cref{eq:optical_moving_redshift} are
\begin{equation}
\widehat k_r=\frac{s_r\sqrt{1-b^2\mathcal{U}(r)}}{\mathcal{F}(r)},\qquad
\widehat k_{\varphi}=-b\sin\iota\cos\chi.
\label{eq:inclined_photon_momenta}
\end{equation}
The initial backward inward leg has $s_r=+1$ for the corresponding future photon; after an exterior turning point, the backward outward leg has $s_r=-1$. On a transmitted backward ray the radius increases in $\mathcal M_2$, so its future-directed photon has $s_r=-1$ there. The sign reversal at the shell is a change of areal chart, not reflection. The same aligned disk coordinates and cumulative angle are used on both sides; the angular targets are not reset at the throat. These signs, together with the specified gas rotation on that side, determine which screen side is Doppler enhanced. A change of screen orientation requires a consistent transformation of the coordinate and momentum conventions.

For circular gas, the frequency ratio can also be written as
\begin{equation}
g_{{\rm circ},i}(r,b,\chi;\iota)=\frac{\sqrt{D(r)/\mathcal F_o}}{1+s_i b\Omega(r)\sin\iota\cos\chi}.
\label{eq:inclined_circular_redshift}
\end{equation}
The angular term produces an approaching--receding brightness difference, whereas the radial term in \cref{eq:optical_moving_redshift} remains necessary for plunging gas. At $\iota=0$, \cref{eq:inclined_first_crossing_angle} gives $\psi_c=\pi/2$, the azimuthal momentum vanishes, and the full construction reduces to \cref{eq:face_on_disk_targets,eq:moving_face_on_circular_weight,eq:moving_face_on_plunge_weight}. The moving face-on and inclined images consequently use identical source and velocity laws.

For inclined viewing, the comparison is defined by the total intensity, its side-resolved contributions, and the radial and azimuthal profiles in \cref{eq:optical_radial_integrated_profile,eq:optical_azimuth_integrated_profile}. The WH and BH share the model-C law and the observer-side rotation. Changes in source amplitude, outer support or counterrotation are separate comparisons at fixed geometry and emission law. The common decomposition in \cref{eq:bilateral_wh_bh_difference} applies at every screen azimuth, so Doppler-enhanced regions can be distinguished from contralateral illumination and BH-source excision. The zero-inclination limit retains the same source supports and observer normalization as the moving face-on construction in Fig.~\ref{fig:face_on_moving_C}; it does not reduce to the static-emitter control.

\begin{figure}[!htbp]
\centering
\includegraphics[width=\textwidth]{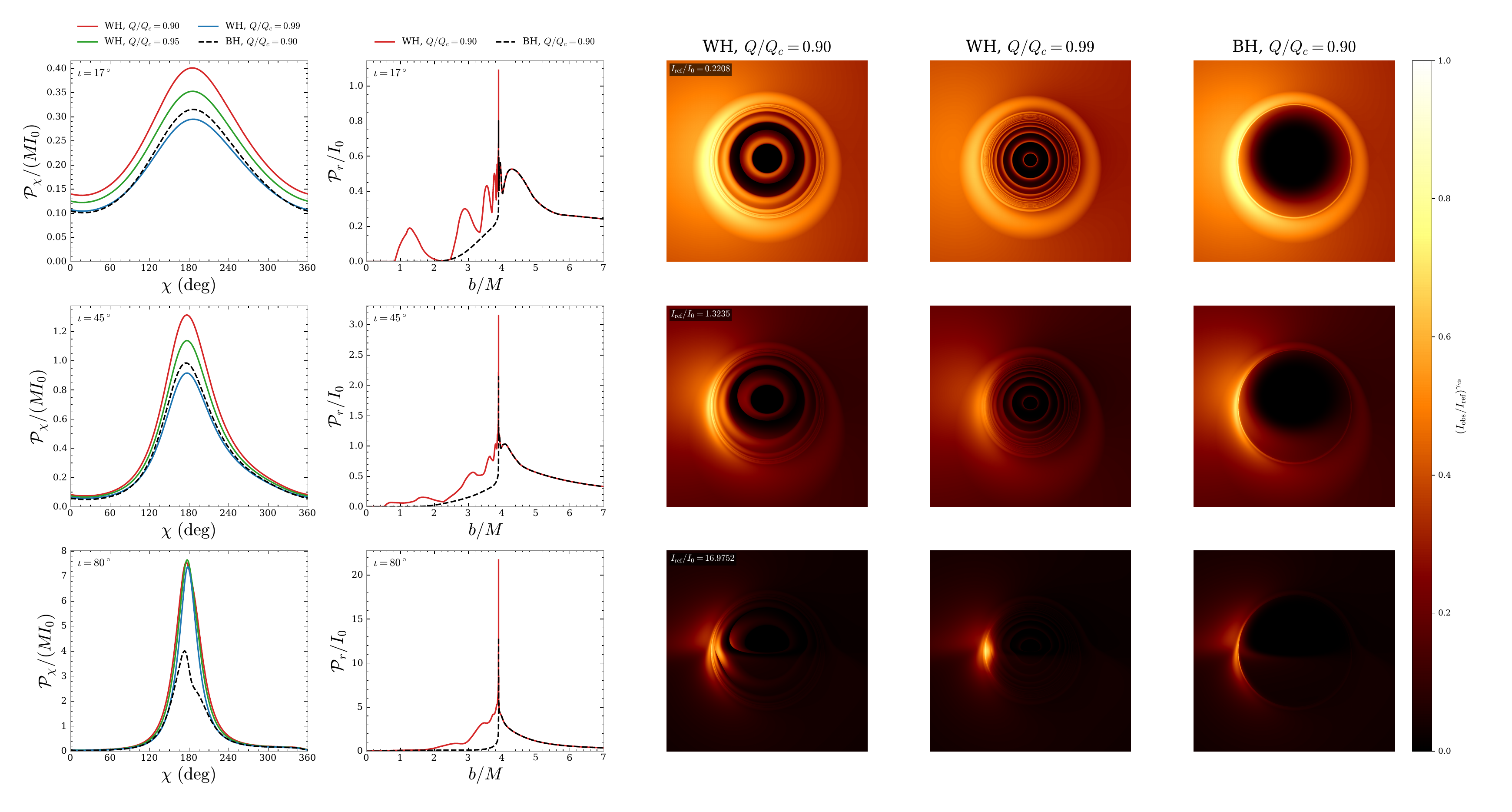}
\caption{Inclined bolometric images of moving model-C emitters at $\nu=0.5$, with $\iota=17^\circ$, $45^\circ$ and $80^\circ$ from top to bottom. The observer is static at $r_o=10^5M$. The first column shows $\mathcal P_\chi/(MI_0)$ from Eq.~\eqref{eq:optical_azimuth_integrated_profile}, with $b_{\max}=7M$, for WHs at $Q/Q_c=0.90$, $0.95$ and $0.99$ and the seed BH at $0.90$. The second shows $\mathcal P_r/I_0$ from Eq.~\eqref{eq:optical_radial_integrated_profile} only for the WH and BH at $0.90$, to keep the radial structure legible. These profiles are not aperture fluxes. The last three columns show WH $0.90$, WH $0.99$ and BH $0.90$, respectively, on $-7\leq X/M,Y/M\leq7$. The source boundary is $10^3M$; each WH has two aligned, equally normalized disks, with $s_1=s_2=+1$ and $a_0=r_h+0.25(r_{\rm ph}-r_h)$. Emitters follow circular geodesics outside the ISCO and its inward separatrix inside. Maps display $(I_{\rm obs}/I_{\rm ref})^{\gamma_{\rm vis}}$ with $\gamma_{\rm vis}=1/2$, distinct from the electromagnetic coupling $\gamma$. One finite-pixel $I_{\rm ref}$ is shared by the three maps within each row, with its value indicated in that row's first map; the references differ between inclinations. All profile curves use the same physical $I_0$.}
\label{fig:inclined_C_nu05_3x5}
\end{figure}

Figure~\ref{fig:inclined_C_nu05_3x5} separates the orientation dependence of the disk image from the geometry of the critical curve. Increasing inclination compresses the projected direct-emission region and produces a progressively stronger azimuthal brightness contrast. The critical curve remains circular at the same radius for a fixed seed, because the background is spherical; the inclined disk images and their bright arcs need not be circular or uniformly illuminated. Equation~\eqref{eq:inclined_first_crossing_angle} changes the crossing radii and accessible intersection sequence with screen azimuth, while Eq.~\eqref{eq:inclined_circular_redshift} enhances circular emission near $\chi=\pi$ for the adopted positive rotation sense and suppresses the opposite side. The fourth power of the frequency ratio in Eq.~\eqref{eq:optical_bolometric_sum} amplifies this contrast. Plunging contributions additionally retain their radial Doppler term and the appropriate future photon sign in each exterior; a single azimuthal correction cannot reproduce the entire image.

The angular profile quantifies the increasing asymmetry on a common physical scale. For the WH at $Q/Q_c=0.90$, its maximum rises from approximately $0.401MI_0$ at $17^\circ$ to $1.315MI_0$ at $45^\circ$ and $7.551MI_0$ at $80^\circ$, while its minimum decreases. The matched BH maxima are approximately $0.315MI_0$, $0.986MI_0$ and $4.008MI_0$. These are maxima of the impact-integrated profile over the specified window, not total luminosities or pointwise critical-ray intensities. At the WH profile's maximum, the opposite-side contribution accounts for approximately $21\%$, $25\%$ and $48\%$ at the three inclinations. Thus identical aligned source prescriptions do not imply an inclination-independent partition of the observed light between the two universes. The changing crossing radii and leg-dependent frequency shifts determine that partition.

The second column isolates the matched $Q/Q_c=0.90$ comparison. Above $b_c/M\simeq3.895$, the WH and BH profiles coincide at all three inclinations: those rays turn outside the throat, and their source intersections and frequency weights are identical. Their differences lie below the common critical impact, where the WH ray can receive transmitted emission. The additional inner structure in the WH maps is consistent with this access to the second disk, but its brightness relative to the BH also includes removal of the BH emitting interval below $a_0$, as expressed by Eq.~\eqref{eq:bilateral_wh_bh_difference}. In the displayed cases, $\mathcal P_\chi$ is larger for the matched WH at every azimuth. This numerical result does not establish a general brightness inequality for arbitrary source supports, amplitudes or velocities. Nor does each radial maximum identify a distinct null orbit: several image orders can accumulate around the same unstable photon barrier.

The omission of $Q/Q_c=0.95$ and $0.99$ from the second column is solely a practical choice of presentation. For this magnetic family, increasing the charge ratio produces a denser sequence of higher-order disk images and radial oscillations, particularly near the critical impact. At $b=0.99b_c$, for example, the accumulated WH angular advance from Eq.~\eqref{eq:optical_angle_integral} increases from approximately $15.52$ to $19.01$ and $34.43$ radians across the three charge ratios. More successive disk-crossing targets can therefore be reached at the same fractional distance from the barrier. Superposing all three WH profiles would obscure both the narrow features and the matched WH--BH comparison. These charge ratios remain physically admitted and represented in the angular profiles, with the $0.99$ case also shown in the maps. Their absence from the radial panels indicates neither a suppressed emission channel nor a restriction of the optical model. The number of distinguishable annuli still depends on resolution and brightness contrast; it is not a finite count of all images in the limiting critical sequence.

Increasing the charge also moves the critical structure inward, from $b_c/M\simeq3.895$ to $3.597$ and $3.170$, consistently with Table~\ref{tab:magnetic_optical_critical_data} and Eq.~\eqref{eq:optical_constrained_charge_response}. This contraction explains the smaller characteristic ring scale of the $0.99$ maps. It accompanies changes in the horizon, ISCO, throat and model-C profile, since the nonlinear magnetic scale obeys Eq.~\eqref{eq:mass_normalized_magnetic_constraint}. More intricate radial structure consequently need not imply a larger integrated brightness. At $17^\circ$ and $45^\circ$, the WH angular maxima decrease with increasing charge across the three samples. At $80^\circ$ they are much closer, and the $0.95$ maximum slightly exceeds the $0.90$ one. This change of ordering reflects the competition between crossing geometry, the intrinsic radial profile and Doppler weighting, rather than a uniform charge enhancement or attenuation. Comparing the $0.99$ WH with the $0.90$ BH changes the seed as well as the topology; only the equal-charge pair isolates the matched throat replacement.

The high-inclination arcs combine projected disk structure with strong frequency weighting, while repeated crossings sustain fine features near the critical circle. Their sampled heights and widths retain the finite-resolution qualification established for the face-on images; the exact critical ray is excluded. Equal colors compare physical intensities between the three maps of one row, but not between different inclinations, whose finite-pixel references differ. Absolute inclination comparisons instead use the profiles normalized by $I_0$. The outer radial components continue to decrease within the displayed window, whereas their broader falloff reflects the source tail discussed with Fig.~\ref{fig:face_on_moving_C}, rather than a new emission component. Finally, the $80^\circ$ result remains a transparent, zero-thickness test-disk calculation. It does not establish the exactly edge-on limit or include absorption, finite-thickness path lengths or a hydrodynamic response of the disks.

\section{Conclusion}\label{sec:conclusion}
Magnetic self-energy links the parameters governing the retained geometry, so the response to charge must be interpreted within a constrained family. For the identical exteriors joined here, changing the charge at fixed asymptotic mass also changes the nonlinear scale. The exact magnetic lapse consequently determines both the available throat domain and the orbital scales entering the emission prescription. The junction still requires negative surface density: the weak and dominant energy conditions fail, while the inner sampled exterior can satisfy the null condition for tangential directions and the strong condition intrinsic to the surface. These distinctions separate the supporting layer from its positive energy magnetic bulk.

The existence of a restoring radial response depends on the constitutive slope at the chosen equilibrium. The calibrated barotropic samples have negative potential curvature, whereas the selected real branches of the generalized and modified cosmic Chaplygin laws admit locally stable sectors. Their analytic endpoints reveal why these sectors need not share the same extent. Near a simple seed horizon, the fractional inverse-density exponents retain a negative divergence that cancels for the unit exponent. Farther outward, competition between the cosmic response and the modified linear pressure contribution controls the leading asymptotic sign and can restrict stability to a finite interval. The constitutive coefficients are fixed during each linear perturbation; the radius scans compare separately calibrated equilibria.

The same flux cancellation that simplifies shell conservation permits an isentropic evolution under the reversible first law at positive temperature. Thermodynamic completions make the remaining entropy-function freedom explicit, thereby separating a radial restoring mode from a thermal stability criterion. The acceleration-based temperature prescription supplies a local scale whose first-order variation includes the shell acceleration and follows the mechanical mode. Its relation to a blueshifted seed-horizon temperature holds in the simple-horizon limit, while the stationary radial null-ray map supplies no particle-creation flux in the adopted vacuum. Thus the mechanical balance and the specified temperature scale have distinct thermodynamic and radiative roles.

Bilateral imaging provides an independent probe of the retained exterior and the source distribution. With minimally coupled Maxwell test radiation and a transparent symmetric junction, both sides have the same critical impact, yet light emitted beyond the throat can reach the observer after a single crossing. For the matched seed black hole and wormhole, rays turning outside the shell produce identical intensity contributions. The inner comparison instead balances the additional transmitted radiation against black-hole emission excluded by the throat cut. This decomposition explains why inner annuli and a smaller dark image region carry information about disk support as well as the spacetime connection.

At fixed source prescription, emitter motion changes the frequency weights while preserving the geometric crossing structure. Circular time dilation suppresses the face-on weight, and the inward continuation from the innermost stable circular orbit adds a radial shift that depends on the photon direction. In the computed profiles, their combination reduces annular intensities relative to the static control. At moderate radii, increasing frequency weights can partly offset the declining intrinsic emission and form a broad shoulder before the inverse-radius falloff resumes. Inclined moving sources develop increasingly uneven brightness around the image, with an orientation-dependent contribution from the opposite disk. Along the sampled magnetic family, larger subcritical charge simultaneously contracts the critical circle and increases the accessible near-critical crossing sequence. These conclusions concern the prescribed finite disks and resolved image structure; the optical throat choices are geometric comparisons rather than selections of mechanically stable equilibria.

A natural next step is to restrict the imaging configurations to the locally stable constitutive sectors and follow how shell motion couples to disk illumination and transit times. A microscopic closure for the surface entropy would complement that dynamical extension. Finite-thickness radiative transfer and the effective characteristics of nonlinear electromagnetic disturbances would then allow the present transparent Maxwell-probe construction to be compared with broader matter and radiation models.

\begin{acknowledgments}
Edson Otoniel acknowledges support from Funda\c{c}\~ao Cearense de Apoio ao Desenvolvimento Cient\'ifico e Tecnol\'ogico (FUNCAP) through grant BP6-0241-00335.01.00/25.
\end{acknowledgments}

\bibliography{ref}

\end{document}